%% file: sample631.tex
\documentclass[twocolumn]{aastex631}

\usepackage{graphicx}
\usepackage{amsmath}
\usepackage{amssymb}
\usepackage{booktabs}
\usepackage{multirow}
\usepackage{makecell}
\usepackage[normalem]{ulem}

\usepackage{dcolumn}
    \newcolumntype{d}[1]{D{.}{.}{#1}}

\newcommand{\kms}{$\rm km~s^{-1}$} 
\newcommand{\lya}{Ly$\alpha$}
\newcommand{\lyb}{Ly$\beta$}

\newcommand{\HI}{\mbox{H\,{\sc i}}}

\newcommand{\OII}{\mbox{O\,{\sc ii}}}

\newcommand{\OVI}{\mbox{O\,{\sc vi}}}

\newcommand{\MgII}{\mbox{Mg\,{\sc ii}}}

\newcommand{\logm}{${\rm log}_{10}(M_{\star}/\rm M_{\odot})$}
\newcommand{\Msun}{$\rm M_{\odot}$}

\newcommand{\kHI}{\mbox{$\kappa_{\tt H\,I}$}}
\newcommand{\kOVI}{\mbox{$\kappa_{\tt O\,VI}$}}

\begin{document}

\title[Probing CGM anisotropies]{MUSEQuBES: Probing Anisotropies in Gas and Metal Distributions in the Circumgalactic Medium} 

\correspondingauthor{Sayak Dutta}
\email{sayak18@iucaa.in}

\author[0009-0000-0797-7365]{Sayak Dutta}
\affiliation{Inter-University Centre for Astronomy \& Astrophysics, Post Bag 04, Pune, India 411007}

\author[0000-0003-3938-8762]{Sowgat Muzahid}
\affiliation{Inter-University Centre for Astronomy \& Astrophysics, Post Bag 04, Pune, India 411007}

\author[0000-0002-0668-5560]{Joop Schaye}
\affiliation{Leiden Observatory, Niels Bohrweg 02, 2333 CA Leiden, Netherlands}

\author[0000-0001-9487-8583]{Sean Johnson}
\affiliation{Department of Astronomy, University of Michigan, 1085 S. University Ave, Ann Arbor, MI 48109, USA}

\author[0000-0002-8505-4678]{Edmund Christian Herenz} 
\affiliation{Inter-University Centre for Astronomy \& Astrophysics, Post Bag 04, Pune, India 411007} 

\author[0000-0002-0873-5744]{Ismael Pessa} 
\affiliation{Leibniz-Institut f\"ur Astrophysik Potsdam (AIP), An der Sternwarte 16, 14482 Potsdam, Germany}

\author[0000-0001-7472-3824]{Ramona Augustin} 
\affiliation{Leibniz-Institut f\"ur Astrophysik Potsdam (AIP), An der Sternwarte 16, 14482 Potsdam, Germany} 

\author[0000-0003-0068-9920]{Nicolas F. Bouch\'e} 
\affiliation{Centre de Recherche Astrophysique de Lyon (CRAL) UMR5574, Univ Lyon1, Ens de Lyon, CNRS, 69230 Saint-Genis-Laval, France} 

\author[0009-0003-3956-4890]{Joey Braspenning} 
\affiliation{Max-Planck-Institut f\"ur Astronomie, K\"onigstuhl 17, D-69117 Heidelberg, Germany} 

\author[0000-0001-5804-1428]{Sebastiano Cantalupo} 
\affiliation{Department of Physics, University of Milan Bicocca, Piazza della Scienza 3, 20126, Milano, Italy}

\author{Sourav Das}
\affiliation{Inter-University Centre for Astronomy \& Astrophysics, Post Bag 04, Pune, India 411007}

\author[0000-0001-5020-9994]{Martin Wendt} 
\affiliation{Institut f\"ur Physik und Astronomie, Universit\"at Potsdam, Karl-Liebknecht-Str. 24/25, 14476 Potsdam, Germany}

\begin{abstract}

\noindent  
We investigate the azimuthal dependence of \HI\ and \OVI-bearing gas in the circumgalactic medium (CGM) of 113 isolated galaxies in the redshift range $0.12<z<0.75$, including 91 new measurements from the MUSE Quasar-fields Blind Emitters Survey (MUSEQuBES). Of these, measurements for 46 galaxies lie within the virial radius ($R_{\rm vir}$), including 36 non-face-on systems for which azimuthal angle ($\phi$) measurements are robust.
The \HI\ covering fraction (\kHI) within $R_{\rm vir}$ of low-mass ($7<$ \logm~$\leq 9$) galaxies, for a threshold column density of ${\rm log}_{10}(N(\HI)/{\rm cm}^{-2})=14.5$, exhibits an enhancement along both the disk plane ($\phi\lesssim20^{\circ}$) and in the polar direction ($\phi\gtrsim70^{\circ}$). In contrast, such a bimodal distribution is not observed for higher mass galaxies ($9<$ \logm~$\leq 11.3$). 
Similarly, the \OVI\ covering fraction (\kOVI), for a threshold of ${\rm log}_{10}(N(\OVI)/{\rm cm}^{-2})=14.0$, shows a tentative enhancement along both the projected major and minor axes for low-mass galaxies. In contrast, \OVI-bearing gas around higher-mass galaxies appears more uniformly distributed, with no significant azimuthal dependence. 
Finally, using the halo circular-velocity-normalized pixel-velocity two-point correlation function (TPCF), we find that \OVI\ absorbers are kinematically narrower along the disk plane compared to the polar directions of the host galaxies with similar stellar mass distributions. 
The observed isotropic distribution of \OVI\ in high-mass halos suggests that its spatial distribution is governed by global halo properties; however, the \OVI\ kinematics retain memory of the site of origin. 
\end{abstract}


\section{Introduction} 
\label{sec:intro}

The evolution of galaxies is governed by complex interactions with their surrounding environment, involving the continuous exchange of matter and energy. Numerical simulations demonstrate that galaxy growth is closely linked to the accretion of intergalactic gas \citep[e.g.,][]{Keres, Voort_2011b}, feedback from galaxies in the form of supernova- and AGN-driven winds \citep[e.g.,][]{Lehnert_2015, Somerville_2015}, recycling of the outflow materials, and the interplay between these processes. While the accretion of cool gas is predicted to occur along the dense, anisotropic intergalactic filaments for lower-mass halos of $M_{\rm halo}\lesssim10^{12}~$\Msun\ \citep[][]{Keres, Voort_2011b}, more massive halos of $M_{\rm halo}\gtrsim10^{12}~$\Msun\ undergoes isotropic accretion of hot, virialized gas \citep[][]{Voort_2011b, Fielding_2017, Hafen_2022}.
The accreted material within the halo is expected to orbit the galaxy, delivering both angular momentum and star-forming fuel, and forming an extended, co-rotating disc \citep[e.g.,][]{Stewart_2011, Stewart_2013, Stewart_2017}. In contrast, outflowing material is predicted to escape preferentially along the projected minor axis, perpendicular to the disk plane, following the path of least resistance \citep[]{Nelson_2019, Mitchell_2020, Peroux_2020, DeFelippis_2020}. In regions where accretion and outflows compete, feedback can inhibit the infall of material from directions perpendicular to the disk, resulting in filamentary accretion that proceeds predominantly co-planar with the galactic disk \citep[e.g.,][]{Stewart_2011, Shen_2012, Voort_12}.

The medium surrounding galaxies, commonly referred to as the circumgalactic medium \citep[CGM; e.g.,][]{Tumlinson_2017, Peroux2_2020, Chen_2024}, is an ideal environment to search for signatures of gas flows predicted by simulations.
However, owing to the very diffuse nature of the gas present in the CGM ($n_{\rm H}\lesssim0.01~{\rm cm}^{-3}$), direct observations of the baryon cycle are only limited to nearby galaxies \citep[e.g., outflows in M82;][]{Castles_1992, Yoshida_2019, Xu_2023}. The direct observation of bipolar outflows in emission in the CGM of galaxies beyond the local universe was only possible with the recent advancement of integral-field-spectroscopy (IFS) with the state-of-the-art integral-field-unit (IFU) detectors such as MUSE \citep[e.g.,][]{Rupke_2019, Zabl_2021, Guo_2023}. 
However, owing to the faintness of the signal, these studies either rely on spectral stacking of many galaxies or are limited to resonant lines (e.g., \MgII) observed in specific types of galaxies, such as starburst galaxies.

The absorption spectra of bright, background sources (such as quasars) have proven to be among the most sensitive probes of the CGM \citep[e.g.,][]{Bergeron_1986, Petitjean_1990, Chen_2009, Tumlinson_2013}. 
Although this method has limitations in distinguishing inflowing from outflowing gas—since the line-of-sight velocity between the galaxy and absorber does not directly probe the radial motion of the gas relative to the galaxy—the additional information provided by the azimuthal angle, defined as the angle between the absorber and the projected major axis of the galaxy, can help differentiate outflows from accreting gas \citep[e.g.,][]{Bouche_2012, Kacprzak_2012}.


\MgII\ absorbers have long been shown to strongly correlate with the orientation of their host galaxies \citep[][]{Bordoloi_2011, Kacprzak_2012, Bouche_2012, Schroetter_2019, Zabl_2019}.
For instance, \citet[]{Kacprzak_2012} identified a bimodal distribution in the azimuthal angles of halo gas traced by \MgII\ absorption, showing a preference for locations near both the projected major and minor axes of star-forming galaxies. In contrast, such a pattern is absent in non–star-forming galaxies. Moreover, the kinematic properties of \MgII\ absorbers are also influenced by galaxy orientation,
 as they tend to be corotating \citep[e.g.,][]{Ho_2017, Zabl_2019}. 

\citet[]{Nielsen_2015} demonstrated that face-on star-forming galaxies for which the quasar sightline is nearly parallel to their minor axes exhibit the largest velocity dispersions, indicating that a portion of \MgII\ absorption likely traces bi-conical outflows.  Nevertheless, more recent studies found no such azimuthal dependence in \MgII\ absorption \citep[see e.g.,][]{Dutta_2020, Huang_2021, Venkat_2025}.
\MgII\ absorbers, however, have been found to co-rotate with their host galaxies in a significant number of studies \citep[][]{Bouche_2016, Martin_2019, Zabl_2019, Lopez_2020}.

Several studies have examined the anisotropies in the gas distribution traced by \HI\ in the CGM of low-redshift galaxies. For instance, \citet[][COS-GASS]{Borthakur_2015} and \citet[][DIISC]{Borthakur_2024} reported no significant correlation between Ly$\alpha$ rest-frame equivalent width ($W_r$) and the azimuthal angle of the host galaxy. In contrast, \citet[]{Beckett_2021} found that galaxies associated with absorbers having $N(\HI)>10^{14}~{\rm cm}^{-2}$ exhibit a bimodal azimuthal angle distribution, with a marked enhancement along both the disk (major axis) and polar (minor axis) directions. Although their sample spans a large range of impact parameters ($D \sim 100 - 1000$ pkpc), they find the trend to be primarily driven by low-impact parameter ($D<500$~pkpc) galaxies.
 On the absorption-selected front of CGM studies, MUSE-ALMA survey \citep[][]{Weng_2023} found marginal evidence of a bimodal azimuthal angle distribution of galaxies associated with strong ($N(\HI)>10^{18}~{\rm cm}^{-2}$) \HI\ absorbers.

A bimodal azimuthal angle distribution has also been reported for highly ionized gas traced by \OVI\ absorption by \citet[]{Kacprzak_2015, Beckett_2021}, which is a probe of metal-rich warm-hot gas, and hence potentially, of galactic winds. They found significant enhancement in the \OVI\ covering fraction between $10^\circ$–$20^\circ$ of the projected major axis and within $30^\circ$ of the projected minor axis. These observations were interpreted as evidence that \OVI\ traces inflowing gas along the major axis and outflowing gas along the minor axis. While \citet[][]{Kacprzak_2019} showed that the kinematics of \OVI\ absorption along the disk plane are not consistent with galaxy rotation in the majority of cases, later studies \citep[][]{Kacprzak_2025, Ho_2026} found evidence for partial and lagging co-rotation of \OVI\ absorbers, with the co-rotation signature weakening at larger galactocentric distances. In contrast to the low-ionization gas phase at smaller radii, which likely traces extended co-rotating structures, the kinematics of the high-ionization phase appear to reflect a more complex interplay among outflows, inflows, and volume-filling warm halos. Hence, \OVI\ kinematics alone may not be a reliable diagnostic of gas flow processes \citep[see also][]{Nielsen_2017, Dutta3_2025}.  

 Recent results from the FOGGIE simulations suggest that \HI\ absorbers preferentially trace the dense cores of small-scale CGM cloudlets, while \OVI\ predominantly arises in their surrounding shells \citep[]{Augustin_2025, Lochhaas_2025}, implying a similar azimuthal variation of \HI\ and \OVI\ absorbers. 
Recent hydrodynamical simulations further predict an increase in gas-phase CGM metallicity along the polar regions compared to the disk plane \citep[see e.g.,][]{Ho_2020, Peroux_2020, Voort_2021, Stern_2021}. On the observational side, however, results remain mixed. While studies such as \citet[]{Pointon_2019} and \citet[]{Sameer_2024} find no significant dependence of CGM metallicity on azimuthal angle, \citet[]{Wendt_2021} report an increase in the gas-phase metallicity inferred from dust depletion ($\rm [Zn/Fe]$) - that is broadly consistent with simulation predictions.

The discussion above highlights a significant inconsistency in the reported azimuthal dependence of CGM properties - such as $W_r$, column density ($N$), and covering fraction of various tracers, including \HI, \MgII, and \OVI, as well as in metallicity measurements. These discrepancies indicate the inherent intricacies of CGM studies, where multiple--often interdependent--parameters such as star formation rate (SFR), stellar mass ($M_{\star}$), impact parameter ($D$), and galactic environment play a critical role in shaping the observed properties \citep[see e.g.,][]{Tum_2011, Johnson_15, Tchernyshyov_2022, Cherrey_2025, Dutta1_2024, MishraN_2024, Dutta2_2025, Dutta4_2025}. As such, isolating the effect of azimuthal angle, which we consider a higher-order dependence, requires careful control over these primary variables.
In this context, we present a focused investigation into the azimuthal angle dependence of \HI\ and \OVI\ absorption in the CGM ($D < R_{\rm vir}$) of isolated galaxies, specifically selected to have no detectable companion galaxy within 500 pkpc and 500 \kms.

 In this work, we primarily use galaxies from the low-$z$ part of the MUSE Quasar-field Blind Emitters Survey \citep[MUSEQuBES;][]{Dutta1_2024, Dutta2_2025, Dutta3_2025} \footnote{For studies of the CGM around $z\approx3.3$ \lya\ emitting galaxies from the MUSEQuBES survey, readers are referred to \citet[]{Muzahid_2020, Muzahid_2021, Banerjee_2023, Banerjee_2025, Banerjee2_2025}.}. The IFU-based galaxy survey without photometric preselection offers a powerful approach to characterizing the galactic environment, especially for the low-mass galaxies. The high signal-to-noise ($S/N$) FUV spectra of the background quasars in the MUSEQuBES fields provide a comprehensive sample of circumgalactic absorption line measurements. Here, we exploit the high-resolution $HST$/ACS imaging of the fields to determine the galaxy morphology and sky orientation. Together, this survey provides a homogeneous galaxy–absorber sample ideally suited for investigating the azimuthal distribution of gas and metals in the CGM.  Drawing from a parent sample of 113 isolated low-$z$ galaxies, we focus on the 36 non-face-on galaxy-absorber pairs located within the virial radius, providing a clean test of azimuthal anisotropy in the gas and metal distributions by minimizing environmental contamination and controlling for key galaxy properties.

This paper is organized as follows. In Section~\ref{sec:data}, we describe the galaxy-absorber sample and the $HST$ imaging. The results are presented in Section~\ref{sec:result}  followed by a discussion in Section~\ref{sec:disc}. Our key findings are summarized in Section~\ref{sec:summary}. Throughout this paper, we adopt a $\Lambda$CDM cosmology with $\Omega_m = 0.3$, $\Omega_{\Lambda}$ = 0.7, and a Hubble constant of $H_0 = 70$ \kms\ ${\rm Mpc}^{-1}$. All distances are in physical units (pkpc, pMpc) unless specified otherwise.

\begin{figure}
    \centering
    \includegraphics[width=1\linewidth]{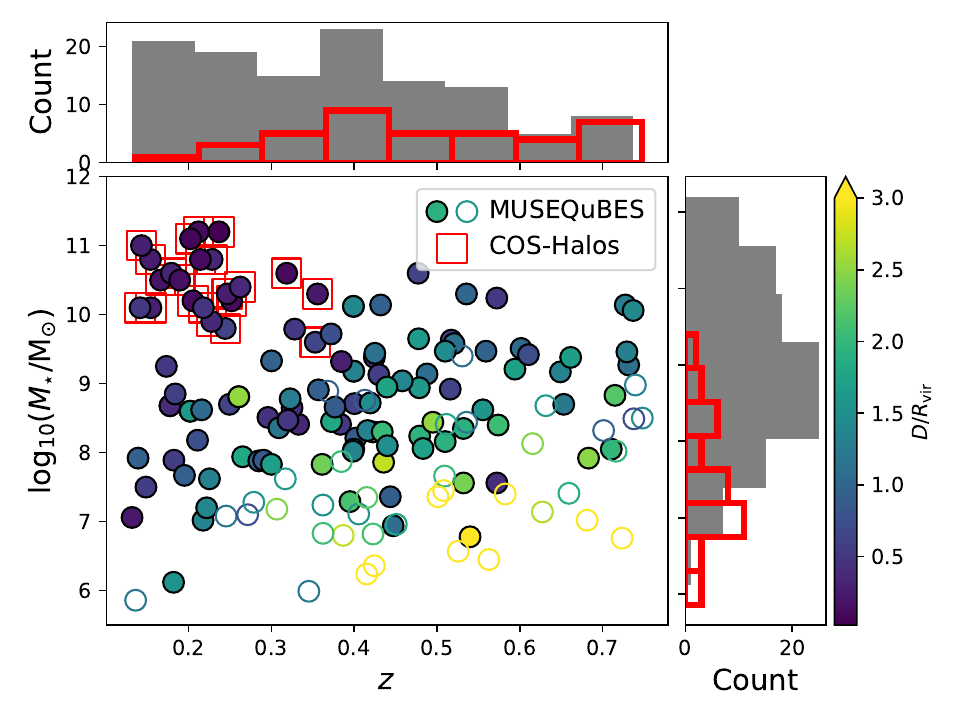}%
    \caption{The stellar masses of the galaxies used in this work are plotted against redshift with solid circles. The COS-halos galaxies are indicated with open red squares. The open circles indicate the MUSEQuBES galaxies discarded from our analysis, which are either undetected or unresolved in $HST$ images (see text). The circles are color-coded by the $D/R_{\rm vir}$ of the galaxies, where the color is saturated at $D/R_{\rm vir}=3$. 
    The filled grey histograms on the top and right side panels show the $z$ and \logm\ distributions, respectively. The red hollow histograms show the same for the discarded galaxies. 
   }  
    \label{fig:galprop} 
\end{figure}

\section{Data} 
\label{sec:data}

\subsection{The galaxy sample} 
\label{sec:gal-sample}

The galaxy sample analyzed in this study is primarily drawn from the low-redshift part of the MUSEQuBES survey \citep{Dutta1_2024, Dutta2_2025, Dutta3_2025}. For a comprehensive overview of the survey design and the methodology for determining galaxy properties, we refer the reader to Section~2 of \citet{Dutta1_2024}. A brief summary is provided below.

The redshifts of these galaxies, detected in the MUSE white-light images, were determined from emission and/or absorption features in the 1D spectra using {\tt MARZ} \citep{Hinton_2016}, with subsequent refinement through a modified version of {\tt PLATEFIT} \citep{Tremonti_2004}. This modified code performs simultaneous Gaussian fitting to all available absorption and emission lines, yielding redshift measurements with uncertainties $<40$~\kms. The SFRs are calculated from {\tt PLATEFIT}--derived H$\alpha$ \citep[]{Kennicutt_1998} or [\OII]  \citep[]{Kewley_2004} line fluxes, corrected to the \citet{Chabrier_2003} initial mass function (IMF) and dust-attenuation via the Balmer decrement when available. The stellar masses of these galaxies are derived through spectral energy distribution (SED) fitting using {\tt FAST} \citep[][]{Kriek_2009}, employing 11 custom pseudo-narrowband images generated from the 1D MUSE spectra after masking the emission lines.
 We use the abundance matching relation from \citet[]{Moster_2013} to derive the corresponding halo mass ($M_{\rm halo}$) and virial radius ($R_{\rm vir}$) from the stellar mass, where $R_{\rm vir}$ is defined as the radius within which the mean enclosed density is 200 times the critical density of the Universe.

We applied a 3D friends-of-friends (FoF) algorithm with a linking velocity of $\pm500$~\kms\ and a transverse linking length of 500~pkpc to identify the isolated galaxies in the MUSEQuBES sample. Additionally, the galaxies are divided into passive (E), star-forming (SF), or unclassified (U) following the approach described in \citet[]{Dutta2_2025}.  Briefly, passive galaxies are defined as those lying more than 3$\sigma$ below the redshift-dependent star-forming main sequence \citep[SFMS;][]{Boogard_18}. Among the remaining galaxies, those with measured star formation rates (SFRs) are classified as star-forming, while those with only $3\sigma$ upper limits on the SFR are categorized as unclassified.

In order to investigate the azimuthal dependence of the circumgalactic gas traced by \HI\ and \OVI, we restrict our galaxy sample to spectroscopically confirmed redshifts in the range $0.12<z<0.75$ within a transverse distance of $300$ pkpc from the background quasars. Additionally, we only select the isolated galaxies such that there are no neighbors within 500 pkpc spatially and 500~\kms\ along the line of sight. This isolation criterion results in a sample of 130 foreground galaxies with $0.12<z<0.75$ and $D<300~$pkpc in the 16 MUSEQuBES fields.

We augmented our sample with 22 galaxies with similar redshift cut from the COS-Halos survey \citep{Werk_2014}, for which high-resolution archival $HST$ imaging is available. By design, the COS-Halos survey pre-selected galaxies without nearby companions having coincident photometric redshifts, thereby favoring systems that are relatively isolated from other $\sim L_*$ galaxies \citep[]{Tumlinson_2013}. While subsequent spectroscopic follow-up revealed the presence of some fainter galaxies at similar redshifts in certain cases \citep[]{Werk_2012}, the COS-Halos sample predominantly consists of isolated galaxies. 

 We note that the isolation condition from MUSEQuBES and COS-Halos are not directly comparable. The MUSEQuBES galaxy sample is complete down to \logm$\lesssim 9$ over the redshift range considered here (Dutta et al., in prep.), whereas the COS-Halos isolation criterion is based on more massive companions. To assess the impact of this difference, we repeated our analysis using a revised MUSEQuBES isolation criterion that excludes only galaxies with neighboring companions of \logm$>9$ within the adopted projected-distance and velocity-separation thresholds.
While the revised criterion of isolation resulted in 13 new `isolated' galaxies, 12 of them are located at $D>R_{\rm vir}$. Hence, this modification does not significantly change our sample.

Fig.~\ref{fig:galprop} shows \logm\ as a function of redshift ($z$) for the full sample of 152 galaxies (130 from MUSEQuBES and 22 from COS-Halos), color-coded by their impact parameter normalized by the virial radius ($D/R_{\rm vir}$).  The COS-halos galaxies are indicated with open red squares. In total 10, 30 and 112 galaxies are categorized as E, U, and SF, respectively.

The stellar masses and redshifts for the COS-Halos galaxies are adopted from \citet[]{Tumlinson_2013}. Galaxies that are not formally detected in the existing $HST$ images (see Section~\ref{sec:hstimages}) are indicated by open circles. The redshift and mass distributions of the sample are shown by the histograms in the side panels of Fig.~\ref{fig:galprop}. The red hollow histograms indicate galaxies excluded from the analysis due to a lack of reliable morphological measurements. It is evident that the discarded galaxies tend to have systematically lower stellar masses, {\bf higher $D/R_{\rm vir}$ }, and lie at higher redshifts.

\subsection{$HST$ Imaging of the quasar fields}
\label{sec:hstimages}

High spatial resolution imaging using space-based instruments is essential to accurately determine the morphological parameters and orientations of the galaxies relative to the background quasar. We obtained $HST$/ACS images for 7 of the 16 MUSEQuBES fields through our program PID: 14660  (PI: Straka). The images of the remaining MUSEQuBES fields were available in the $HST$ public archive. In addition, we make use of $HST$ imaging for 15 quasar fields from the COS-Halos survey, covering all the 22 galaxies. We have used $HST$/ACS (F814W) or $HST$/WFC3 (F160W, F140W and F390W) images of the galaxies. The details of the $HST$ observations are summarized in Table~\ref{tab:hst_obs} of Appendix \ref{sec:hst-summary}.  All of the data presented in this article were obtained from the Mikulski Archive for Space Telescopes (MAST) at the Space Telescope Science Institute. The specific observations analyzed can be accessed via \dataset[10.17909/38m8-6j96]{ https://doi.org/10.17909/38m8-6j96}.

Through visual inspection, we excluded 39 galaxies from our analysis that are either unresolved or not formally detected in $HST$ data. This is further discussed in Appendix \ref{sec:disc-sample}. This results in a final sample of 113 galaxies (91 MUSEQuBES + 22 COS-Halos). Among these, five MUSEQuBES galaxies are probed by two distinct quasars at different impact parameters and azimuthal angles, yielding a total of 118 quasar-galaxy pairs.

Fig.~\ref{fig:inc_sample} shows the $HST$ cutouts for the final sample of 118 quasar–galaxy pairs arranged by their stellar masses. Galaxies from the MUSEQuBES survey are outlined in green, while those from COS-Halos are outlined in red. Complementing the predominantly low-mass MUSEQuBES galaxies, the COS-Halos sample contributes relatively more massive systems. Each cutout spans 20 times the effective area, $\pi R_e^2 {\rm cos}~i$, of the galaxies, 
where $R_{e}$ and $i$ are the most-probable effective radius and inclination angle derived from the posterior distribution using the Bayesian {\sc Galfit} wrapper as described in the section below. Face-on galaxies, defined as galaxies with most probable $i<35^{\circ}$, are additionally highlighted with white boxes around the most probable $i$ values annotated in the bottom right corner of the panels.

\begin{figure*}
    \centering
    \includegraphics[width=\textwidth]{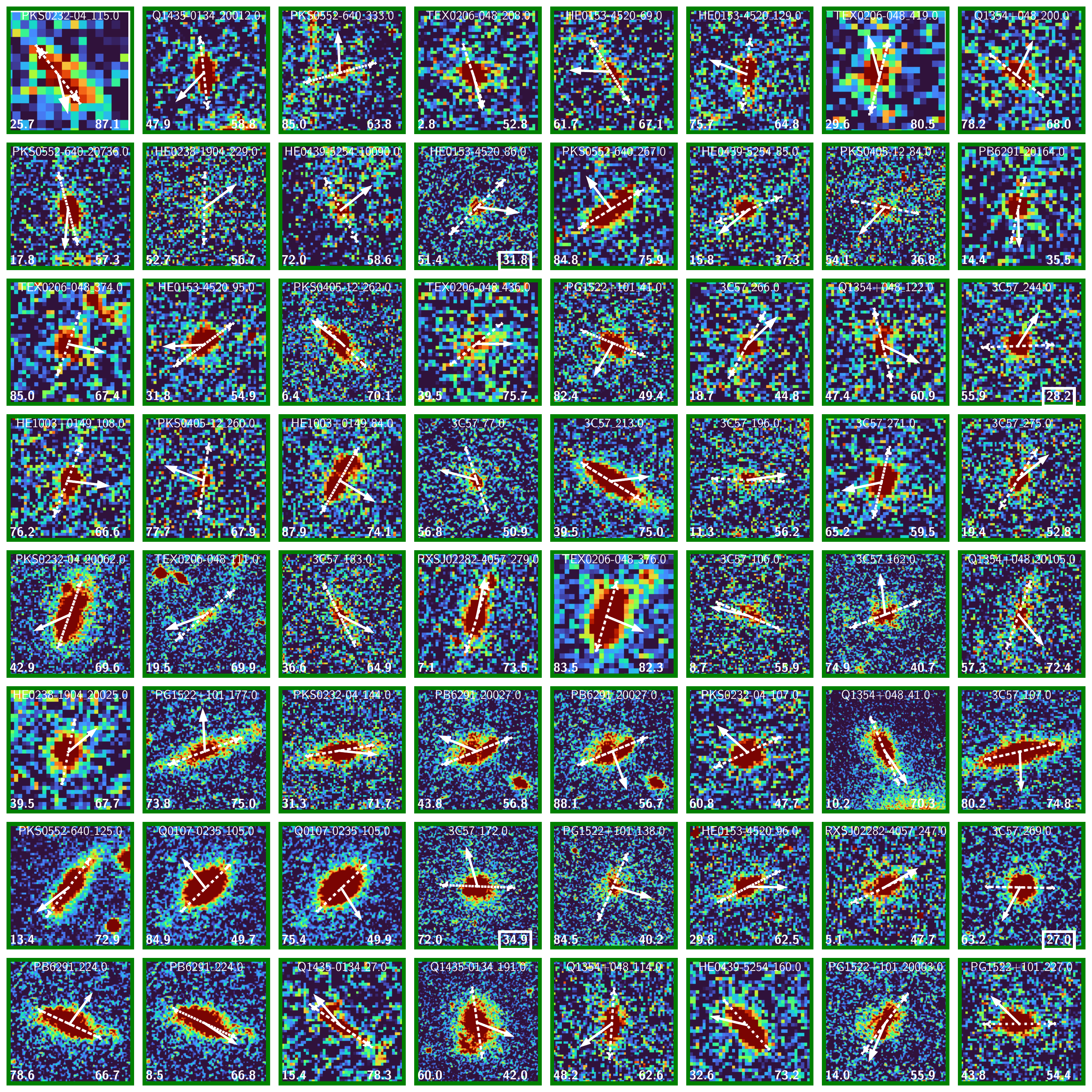}
    \caption{Cutouts of 113 galaxies, sorted by increasing stellar mass, constituting the 118 quasar-galaxy pairs used in this work. Note that 5 galaxies appear more than once since they are probed by multiple quasars.  Each cutout spans 20 times the effective area of the galaxies. Green borders indicate galaxies from the MUSEQuBES sample, while red borders denote the 22 galaxies from the COS-Halos sample. The white solid and dashed arrows in each panel indicate the direction of the quasar location and projected major axis, respectively. The numbers in the lower left and right in each panel represent the most probable $\phi$ and $i$, respectively, with the white boxes highlighting the face-on galaxies.}     
    \label{fig:inc_sample}
\end{figure*}



\begin{figure*}
\addtocounter{figure}{-1}
    \centering
    \includegraphics[width=1\linewidth]{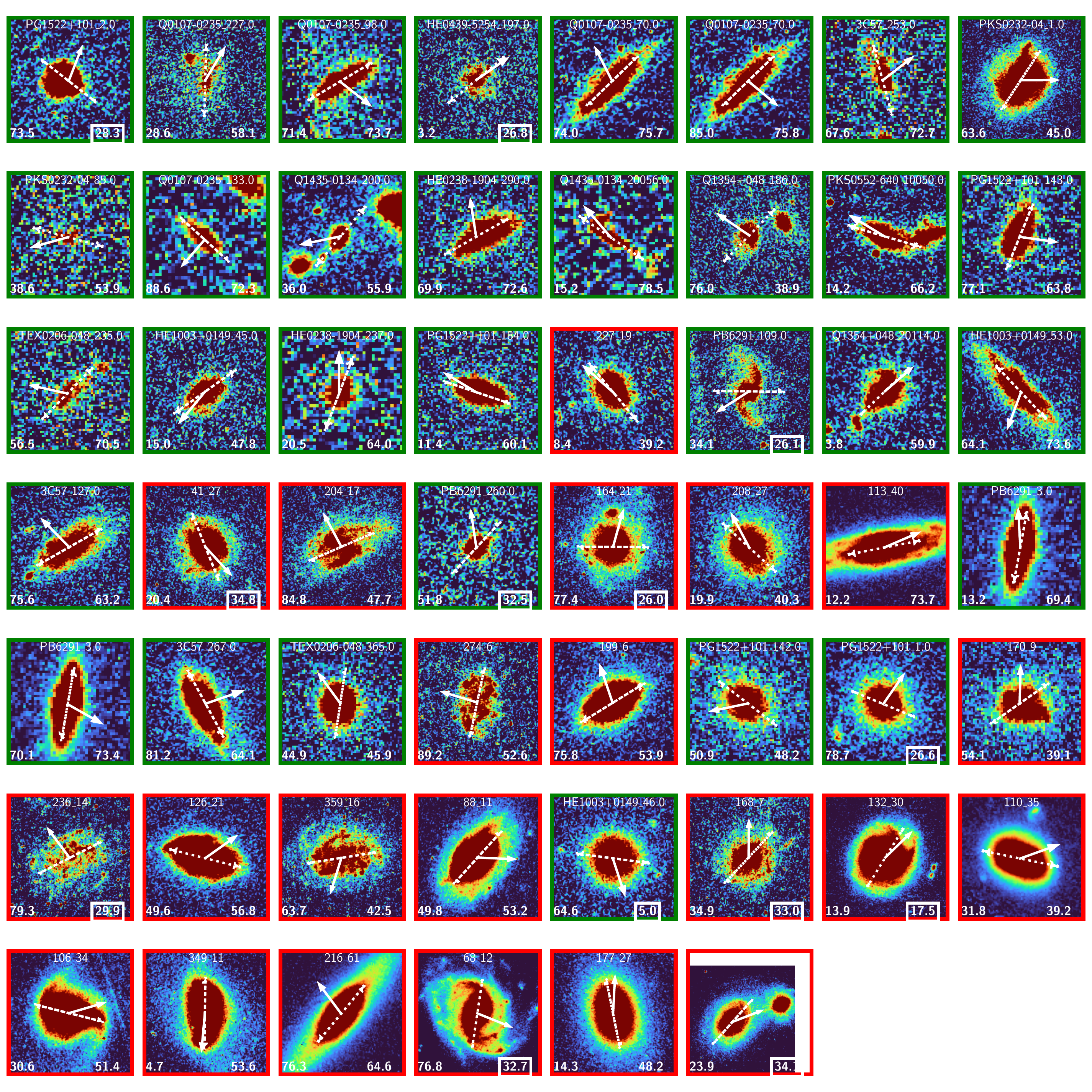}
    \caption{-continued.}   
    \label{fig:inc_sample2}
\end{figure*}

\subsection{Bayesian galaxy modeling with GALFIT} 
\label{sec:galfit}

In this section, we describe the methodology used to extract morphological parameters of the galaxies in our sample from $HST$ imaging.

In this work, we use {\sc Galfit} \citep[][]{Peng_2002, Peng_2010} to fit a single-component Sérsic profile \citep[]{Sersic_1963} to the two-dimensional surface brightness distribution of each galaxy. {\sc Galfit} uses $\chi^2-$minimization to constrain the best-fit effective radius ($R_e$), 
axis-ratio ($b/a$),
position angle ($PA$) of the stellar disk, the position of the galaxy center ($R_0$), AB apparent magnitude ($m$), and the Sérsic index ($n$).
The inclination angle\footnote{We adopt the standard convention for the inclination angle, where $i = 0^{\circ}$ ($i = 90^{\circ}$) corresponds to a face-on (edge-on) orientation.} ($i$) is obtained from $b/a$ as $i = {\rm cos}^{-1}(b/a)$. 
The uncertainties in the best-fit parameters are derived from the weight image (also known as $\sigma-$image), i.e., the standard deviation of counts at each pixel based on Poisson statistics, internally generated by {\sc Galfit}. It is well-known that the best-fit parameters returned by {\sc Galfit} can be sensitive to the choice of initial guess values. Furthermore, the formal uncertainties reported by {\sc Galfit} are often underestimated \citep[]{Peng_2002}. This is particularly crucial for the low-mass galaxies, where the relative faintness can limit the robustness of the morphological parameters.

Alternatively, adopting a Bayesian approach in place of traditional $\chi^2$-minimization allows one to sample the posterior distributions of the morphological parameters, yielding more robust and realistic estimates of their uncertainties. To achieve this, we develop a custom Bayesian wrapper around the traditional {\sc Galfit}, enabling us to sample the posterior distributions of the fitting parameters and derive more reliable uncertainty estimates. Below, we describe the procedure step-by-step:   
(1) First, we select $201\times201$-pixel cutouts of the galaxies from the $HST$ images to perform the fitting.
(2) 
To account for the point-spread function (PSF), we modeled it at the relevant detector chip location using {\tt TinyTim} \citep{Krist_2011}.
(3) Rather than relying on the built-in $\chi^2$-minimization to determine the best-fit parameters, we sample the parameter space using 1000 walkers via the Python package {\tt emcee} \citep[][]{Foreman-Mackey_2013} to obtain the posterior log-likelihood distribution. For each sampled parameter set, we compute the residual image, $\mathcal{R}$, by subtracting the {\sc Galfit} model from the observed data. Using the $\sigma-$image returned by {\sc Galfit}, we calculate the likelihood as: 


\begin{equation}
    \mathcal{L}(R_0,PA,m,R_e,n,b/a ) \propto \prod_{i,j} \frac{1}{\sqrt{2\pi\sigma_{ij}^2}} e^{-\mathcal{R}_{ij}^2/2\sigma_{ij}^2}~.  
\end{equation}


We use Gaussian priors for the galaxy positions ($R_0$) using the RA and Dec values from the MUSEQuBES and COS-Halos catalogs, where the width of the Gaussian priors are set to 2 pixels. Flat priors\footnote{$PA\in (-90, 90)$, $m\in (15,28)$, $R_e\in (1,30)$, $n\in(0.1,8)$, $b/a\in(0,1)$.} were used for the remaining parameters. The most probable values and the 68\% credible intervals from the posterior distributions are used as best-fit solutions and the associated uncertainties. We show an example of our fitting procedure in Appendix \ref{sec:fit-proc}.

We convert the best-fit position angle ($PA$) of the galaxy, as returned by {\sc Galfit}, into the azimuthal angle, $\phi$, which quantifies the angle between the projected major axis of the galaxy and the line in the plane of the sky connecting the galaxy center to the background quasar. Specifically, the position angle, $PA_{1,2}$, of the line joining the galaxy and the quasar center is given by: 
\begin{equation}
    \tan(PA_{1,2}) = \frac{\sin(\alpha_1-\alpha_2)}{\cos\delta_2\tan\delta_1-\sin\delta_2\cos(\alpha_1-\alpha_2)}~,
\end{equation}
where ($\alpha_1$, $\delta_1$) and ($\alpha_2,\delta_2$) are the right ascension and declination of the galaxy and quasar, respectively. The angle $\phi$ is then defined as $\phi = |PA - PA_{1,2}|$. We restrict the azimuthal angle $\phi$ to the range $0^{\circ}$ to $90^{\circ}$ by defining $\phi = \min(\phi,~180^{\circ} - \phi)$. This convention ensures that $\phi = 0^{\circ}$ corresponds to the projected major axis, while $\phi = 90^{\circ}$ corresponds to the projected minor axis.

\begin{figure}
    \centering
    \includegraphics[width=1\linewidth]{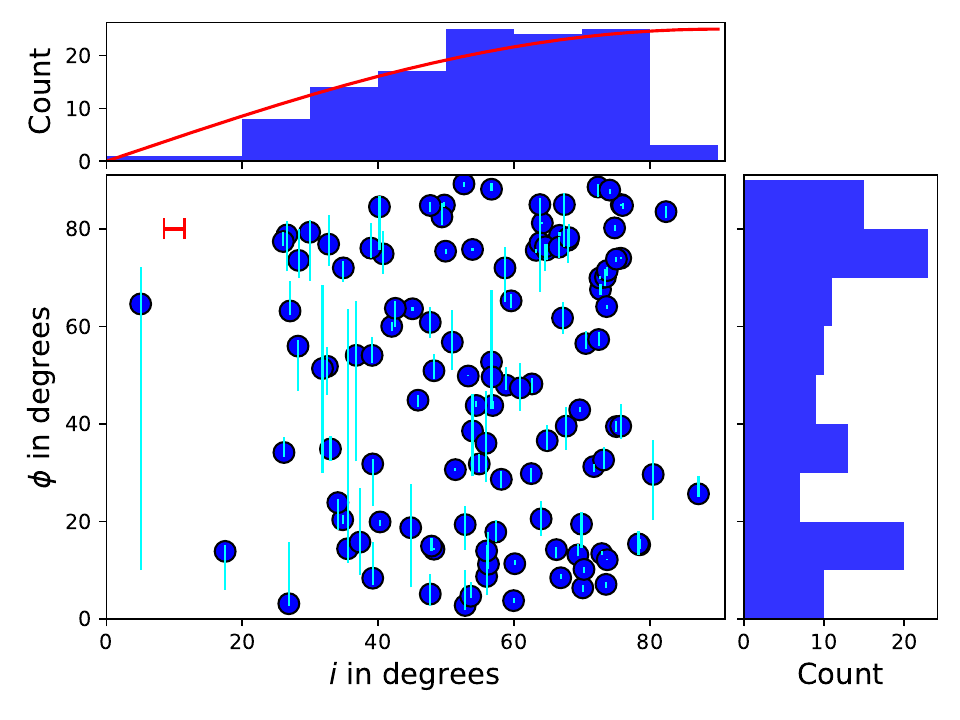}    
    \caption{The most probable azimuthal angle ($\phi$) plotted against the most probable inclination ($i$) for the galaxies used in this work. The y error bars indicate the 68\% credible intervals obtained from the posterior $PA$ distributions. The median 68\% confidence interval on $i$ is shown with the red horizontal bar on the top left. The side panels on the top and right show the most probable $i$ and $\phi$ distributions. 
    The solid red line represents a scaled ${\rm sin} ~i$ function expected for randomly oriented disk galaxies.}  
    \label{fig:galprop_morph}
\end{figure}

In Fig.~\ref{fig:galprop_morph}, the most probable $\phi$ is plotted against the most probable $i$ for the galaxies in our sample. The y and x error bars indicate the 68\% credible intervals of $\phi$ and $i$  propagated from the $PA$ and $b/a$ posterior distributions, respectively. The side panels on top and right show the most probable $i$ and $\phi$ distribution of galaxies, respectively. 
The histogram of $i$ follows the well-known $\sin i$ distribution as indicated by the solid red curve. The distribution of most probable $R_e$, $n$, and $PA$ are shown in Appendix \ref{sec:best-fit-prop} (Fig.~\ref{fig:prop_dist}).

\begin{figure*}
    \centering
    \includegraphics[width=1\linewidth]{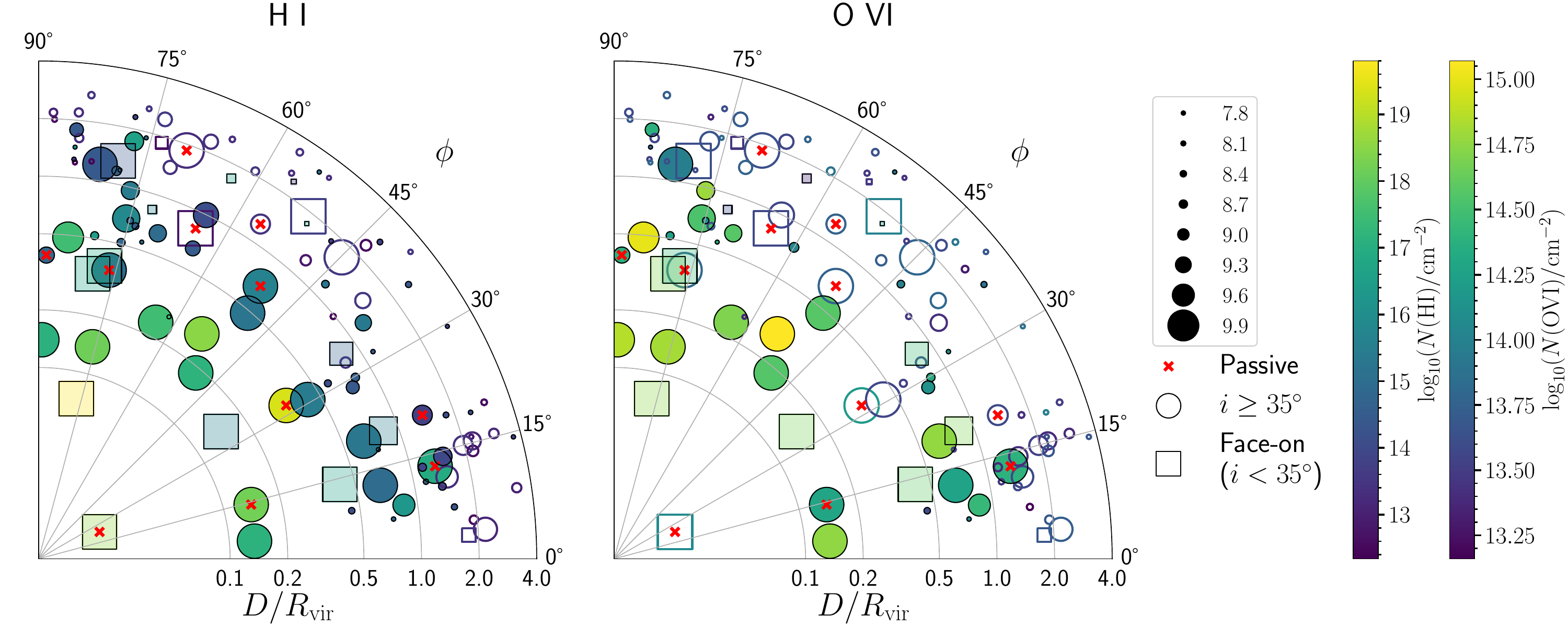}  
    \caption{{\tt Left:} \HI\ column density (color) shown as a function of normalized impact parameter, $D/R_{\rm vir}$, along the radial direction and the azimuthal angle, $\phi$ along the polar direction. The filled and open symbols represent detections and $3\sigma$ upper limits, respectively. The circles and squares represent measurements for edge-on (most probable $i\geq 35^{\circ}$) and face-on (most probable $i< 35^{\circ}$) galaxies, respectively. Face-on galaxies are not considered in subsequent analyses. The area of the points scales with stellar mass, as shown (only for edge-on galaxies) in the legend.
    {\tt Right:} Same as {\tt left} but for \OVI. In both panels, one measurement at $D/R_{\rm vir}>4$ is not shown.}     
    \label{fig:N_phi_all}
\end{figure*}

\subsection{The \HI\ and \OVI\ absorber catalogs}  
\label{sec:abs-cat}

For the MUSEQuBES galaxies, the \HI\ absorber catalog is presented in \citet[]{Dutta4_2025}, while the \OVI\ absorber catalog is provided in  \citet{Dutta2_2025}. Here, we adopt those catalogs for our analysis. 
Briefly, the \HI\ absorption catalog was constructed through a systematic search for Ly$\alpha$ and Ly$\beta$ lines within $\pm600$~\kms\ of MUSEQuBES galaxy redshifts. We derived robust \HI\ column densities for the detected absorbers using multi-component Voigt-profile fitting, simultaneously incorporating all available Lyman series transitions that are free from major contamination in the fit. The individual Voigt components are sorted by velocity and then grouped into \HI\ {\it systems} using a 1D FoF algorithm with a linking velocity of 300 \kms. 
The total \HI\ column density of the system, $N(\HI) \equiv \sum N(\HI)_{\rm comp}$, is then associated with a galaxy if any component of the system falls within $\pm300$~\kms\ of the galaxy's redshift.

For the \OVI\ transitions, a `galaxy-blind' catalog was constructed for the 16 MUSEQuBES sightlines \citep[see][for details]{Dutta2_2025}. The same association method used for \HI\ was then applied to link \OVI\ {\it system} column densities ($N(\OVI)$) with the MUSEQuBES galaxies. In cases of non-detections, we estimate $3\sigma$ upper limits on the \HI\ and \OVI\ column densities from the standard deviation of the normalized continuum flux over a velocity window of $\pm60$~\kms. This window corresponds to approximately twice the median Doppler $b$-parameter of the detected \HI\ and \OVI\ absorption components.

For the COS-Halos galaxies, we obtained the \HI\ absorption measurements (detected column densities and $3\sigma$ upper limits) from \citet[]{Tumlinson_2013}. For the saturated absorbers associated with the COS-Halos galaxies, we have obtained the \HI\ column densities reported in \citet[]{Prochaska_2017}, who used both low and medium resolution $HST$/COS spectra, using G140L/1280 and G130M/1222 gratings respectively, to constrain the \HI\ column densities from the Lyman limit breaks.
The total \OVI\ column densities associated with the COS-Halos galaxies (along with 3$\sigma$ upper limits measured over $\pm50$ \kms\ for non-detections) were obtained from \citet[]{Tumlinson_2011}. In addition, we utilized the individual component column densities and Doppler $b$-parameters reported in \citet[]{Werk_2013} for these galaxies. Note that the detected \HI\ and \OVI\ absorbers associated with the COS-Halos galaxies are measured within a velocity window of $\pm300$~\kms, consistent with the velocity window adopted in our MUSEQuBES survey. The \HI\ and \OVI\ absorption line measurements are tabulated in Appendix \ref{sec:full_sample_table}.

\section{Results} 
\label{sec:result}

The polar plots in the left and right panels of Fig.~\ref{fig:N_phi_all} show the color-coded \HI\ and \OVI\ column densities, respectively, as a function of $D/R_{\rm vir}$ (radial direction) and $\phi$ (angular direction). The points with red crosses indicate the passive galaxies. Among the remaining galaxies, 8 are of type `U', i.e., unclassified, while the rest are star-forming. The areas of the circles are proportional to the stellar mass of the host galaxies. The dotted squares in both panels indicate the 16 face-on galaxies (most probable $i<35^{\circ}$), which are not included in our analyses due to the large uncertainties in their position angle measurements.

It is evident from the left panel of Fig.~\ref{fig:N_phi_all} that the strong \HI\ absorbers are predominantly detected in the inner CGM ($D/R_{\rm vir} \lesssim 0.5$) of relatively massive galaxies. In contrast, the low-mass MUSEQuBES galaxies, primarily probing larger impact parameters ($D/R_{\rm vir} \gtrsim 0.5$), tend to exhibit weaker \HI\ absorption. We caution that the present sample does not include low-mass galaxies with $D \lesssim 0.5R_{\rm vir}$, constraining our ability to probe the inner CGM of low-mass galaxies.
A handful of low-mass galaxies showing relatively strong \HI\ absorption ($N(\HI) \gtrsim 10^{14.5}~{\rm cm}^{-2}$) are mostly probed along the projected major ($\phi \lesssim 20^{\circ}$) or minor ($\phi \gtrsim 70^{\circ}$) axes. \HI\ non-detections are predominantly found beyond the virial radius ($D > R_{\rm vir}$). Overall, we do not observe a significant trend between \HI\ column density and azimuthal angle.

 Contrary to \HI\ where non-detections are primarily located beyond $R_{\rm vir}$, the non-detections of \OVI\ are prevalent even within $R_{\rm vir}$, both for very massive and low-mass galaxies. In the case of the most massive galaxies, this lack of \OVI\ detection is primarily driven by their passive nature \citep[sSFR$\lesssim10^{-11}~{\rm yr}^{-1}$, see e.g., ][]{Tchernyshyov_2023, Dutta2_2025}.
For the lower mass galaxies, \OVI\ is almost always detected within $R_{\rm vir}$ along the major ($\phi\lesssim20^{\circ}$) and minor ($\phi\gtrsim70^{\circ}$) axes, while the non-detections within $R_{\rm vir}$ are mostly at intermediate $\phi$ ($\approx 20^{\circ}-70^{\circ}$). Outside $R_{\rm vir}$, galaxies generally do not exhibit detectable \OVI\ absorption at any $\phi$.

In the next section, we investigate potential trends between the \HI\ and \OVI\ covering fractions and the azimuthal angles of the associated galaxies, in order to probe possible anisotropies in the distribution of gas and metals in and around galaxies.

\subsection{Variation of covering fraction with $\phi$}
\label{sec:sec3.1}

In this section we will investigate the variation of \HI\ and \OVI\ covering fraction as a function of the azimuthal angle. In order to properly account for the uncertainties in $\phi$ measurements, we followed an approach similar to \citet{Kacprzak_2015}. Briefly, for each galaxy, we use the posterior $\phi$ distribution from our Bayesian {\sc Galfit}  analysis to generate a probability density function (PDF) on a $\phi$ grid of 0$^{\circ}$-90$^{\circ}$ in steps of 0.09$^{\circ}$. We then compute the total $\phi$ PDF across galaxy subsamples. Finally, we bin this total $\phi$ PDF into $18^{\circ}$ or $30^{\circ}$--wide azimuthal angle bins for analysis. This procedure naturally incorporates measurement uncertainties, with galaxies having tighter posterior distributions in $\phi$ contributing more strongly to azimuthal bins nearer their most probable values.

The covering fraction ($\kappa$) is subsequently obtained by dividing the number of galaxies with detected absorption, $n_1$,
 by the total number of galaxies, $n_1+n_2$, where $n_2$ is the number of galaxies with non-detections.
A galaxy is considered to have detected absorption if its CGM absorption exceeds a threshold column density.
The median spectral sensitivity of the quasar sample used in this work ensures that the $3\sigma$ upper limits on $N(\HI)$ and $N(\OVI)$ are systematically lower than the minimum column density thresholds adopted in this study.
Both $n_1$ and $n_2$ in a given $\phi-$bin are obtained by multiplying the binned $\phi$ PDF with the bin-width of $18^{\circ}$ or $30^{\circ}$. The uncertainties in $\kappa$ are reported using the 68\% Wilson score intervals, as well as 68\% confidence intervals derived from 10,000 bootstrap realizations of the galaxy sample. 
\begin{figure*}
    \centering
    \vskip-0.3cm 
    \includegraphics[width=0.33\linewidth]{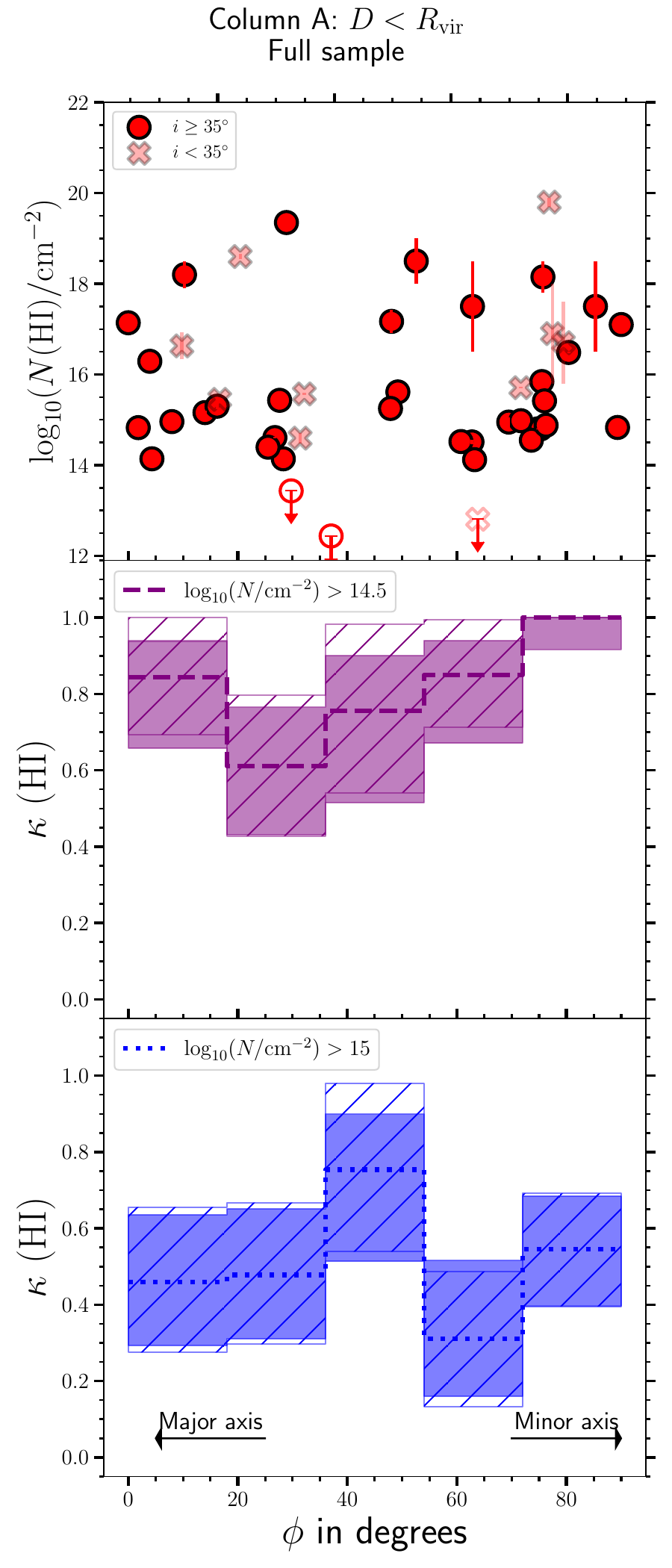}%
    \includegraphics[width=0.33\linewidth]{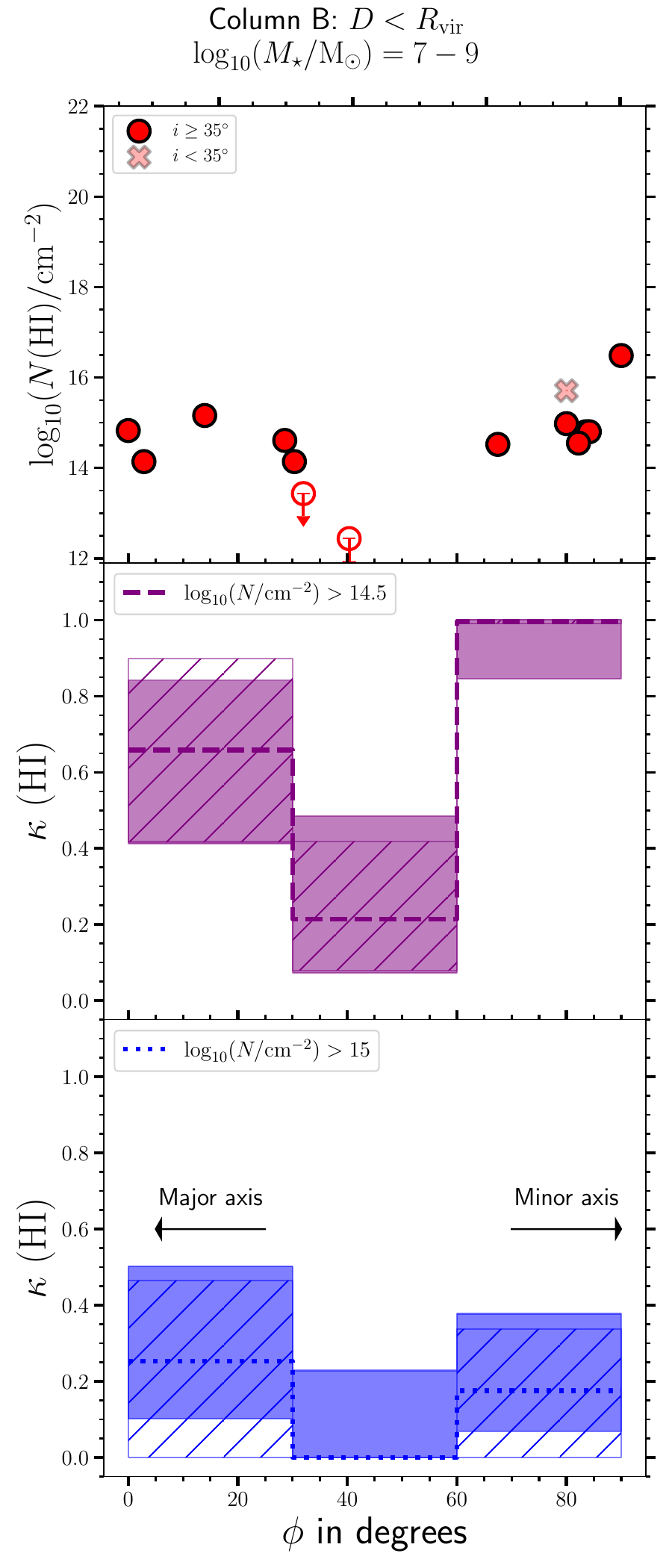}%
    \includegraphics[width=0.33\linewidth]{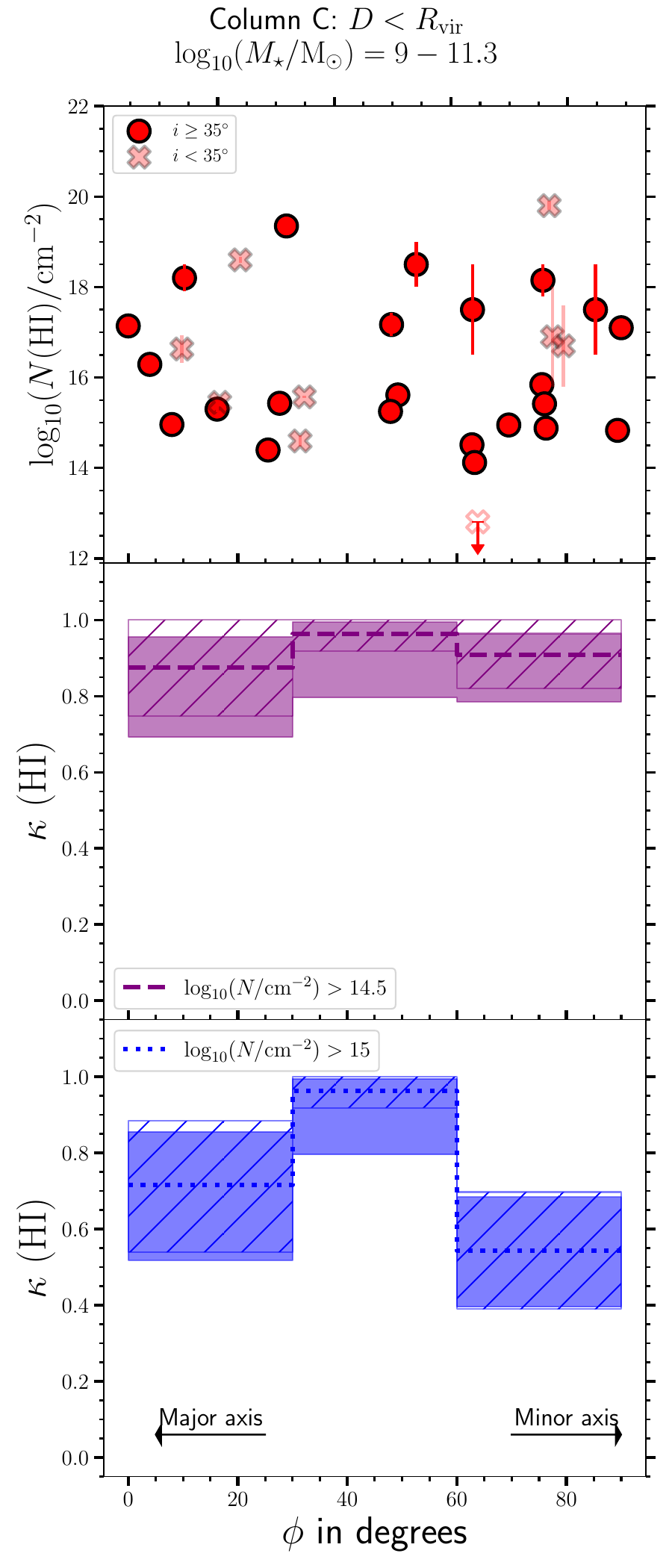}      
    \caption{{\tt Column A, Top:} The red solid and open circles show the detected $N(\HI)$ and $3\sigma$ upper limits for non-detections, respectively, plotted against the most probable $\phi$ for galaxies with $D<R_{\rm vir}$. The red crosses indicate the discarded face-on galaxies with most probable $i<35^{\circ}$. {\tt Column A, Middle:} The \HI\ covering fraction (\kHI) for a threshold $N(\HI)=10^{14.5}~{\rm cm}^{-2}$ plotted against $\phi$ for the galaxy sample shown in the left panel (discarding the face-on galaxies). The shaded and hatched regions represent the 68\% Wilson score interval and 68\% confidence interval obtained from 10000 bootstrap realizations of the galaxy sample, respectively. {\tt Column A, Bottom:} Same as {\tt middle}, but for a threshold $N(\HI)=10^{15}~{\rm cm}^{-2}$. No significant $\phi$ dependence is seen in any of these panels. {\tt Columns B} and {\tt C} is same as {\tt Column A} but for dwarf (\logm$\approx7-9$) and massive (\logm$\approx 9-11.3$) galaxies, respectively. A marginal reduction in \kHI\ is seen for the intermediate $\phi$ bin for the dwarf galaxies ({\tt Column B}). However, the \kHI\ for more massive galaxies does not exhibit any significant azimuthal dependence ({\tt Column C}). }
    \label{fig:hi_all}
\end{figure*}

\subsubsection{Azimuthal dependence of \HI\ in the CGM}

In the left panel of {\tt Column A} in Fig.~\ref{fig:hi_all}, individual $N(\HI)$ measurements within the virial radius of 46 galaxies are plotted against the most probable $\phi$ with solid and hollow red circles indicating the detections and 3$\sigma$ upper limits for non-detections, respectively. The cross symbols indicate face-on galaxies with most probable $i<35^{\circ}$. No significant variation of the measured $N(\HI)$ with $\phi$ is observed within $R_{\rm vir}$; however, it is noteworthy that non-detections are predominantly concentrated at intermediate azimuthal angles.
In the middle and bottom panels of {\tt Column A} in Fig.~\ref{fig:hi_all}, we show the \HI\ covering fraction (\kHI) against $\phi$ in 18$^{\circ}$ bins, for the 36 galaxies with $D<R_{\rm vir}$ and $i>35^{\circ}$ for a threshold $N(\HI)$ of $10^{14.5}~{\rm cm}^{-2}$ and $10^{15}~{\rm cm}^{-2}$, respectively. While the 10 face-on galaxies ($i<35^{\circ}$) were formally excluded from all covering fraction analyses, we confirmed that our results 
are robust with respect to this inclination angle cut.
The \kHI\ is marginally enhanced along the major ($\kHI=0.84^{+0.16}_{-0.14}$ for $\phi\leq 18^{\circ}$) and minor ($\kHI=1.00^{+0.00}_{-0.08}$ for $\phi\geq72^{\circ}$) axes compared to intermediate $\phi$ ($\kHI=0.61^{+0.18}_{-0.18}$ for $\phi=18^{\circ}-36^{\circ}$) for the threshold of  $10^{14.5}~{\rm cm}^{-2}$. However, no significant variation of \kHI\ with $\phi$ is observed for the full sample for the threshold of $N(\HI)=10^{15}~{\rm cm}^{-2}$.

 Next, we divide our galaxy sample - spanning a broad dynamic range in stellar mass - into two $M_{\star}$ bins: \logm\ $=7$–$9$ (median 8.4) and \logm\ $=9$–$11.3$ (median 10.1). 
 {\tt Columns B} and {\tt C} of Fig.~\ref{fig:hi_all} present the same analysis as panel {\tt Column A}, but separately for the 13 low-mass, and 23 massive galaxies, respectively.
Note that we have discarded one dwarf and 9 massive galaxies with $i<35^{\circ}$ in this analysis.  
Here we used a larger $\phi$-bin of 30$^{\circ}$ to ensure a statistically significant number of galaxies in each $\phi$ bin.

The \kHI\ is enhanced along both the projected major ($\kHI=0.66_{-0.25}^{+0.18}$ for $\phi\leq30^{\circ}$) and minor ($\kHI=1.00_{-0.16}^{+0.00}$ for $\phi\geq60^{\circ}$) axes compared to intermediate $\phi=30^{\circ}-60^{\circ}$ ($\kHI=0.21_{-0.14}^{+0.27}$) for galaxies with \logm\ $=7$–$9$, for the lower column density threshold of $N(\HI)=10^{14.5}~{\rm cm}^{-2}$. The enhancement is not significant for the higher threshold of $10^{15}~{\rm cm}^{-2}$. However, due to the modest sample size of dwarf galaxies, this result should be interpreted with caution.
In contrast, the \kHI\ for the high-mass galaxies with \logm\ $=9$–$11.3$ does not show any significant azimuthal dependence for either of the thresholds. We note the 16-50-84 percentile of $D/R_{\rm vir}$ distribution of the low- and high-mass samples are 0.5-0.7-0.8 and 0.1-0.4-0.8, respectively. In order to mitigate any systematics caused by the inner CGM measurements of high-mass galaxies that are unavailable to low-mass counterparts, we split the high-mass galaxy sample into two bins with $D/R_{\rm vir}\leq 0.5$ and $D/R_{\rm vir}>0.5$. \kHI\ does not exhibit any significant azimuthal variation for the high-mass galaxies in either of the $D/R_{\rm vir}$ bins (Appendix, left panel of Fig.~\ref{fig:app:2dn_bins}). 
Furthermore, we find no significant $\phi$-dependence of \kHI\ for galaxies with $D > R_{\rm vir}$ across any of the adopted $N(\HI)$ thresholds, in either the low- or high-mass galaxy bins (Appendix, Fig.~\ref{fig:app:out_vir}). 

Finally, we have examined the $\phi-$dependence of \kHI\ for galaxies selected by impact parameter (not shown). For galaxies with $D<50$ pkpc, neither the full sample nor the high-mass subsample shows any indication of bimodality for either threshold $N(\HI)$. However, the low-mass galaxies (with a modest sample of 8 galaxies) exhibit enhanced \kHI\ along the major and minor axes for the threshold $N(\HI)=10^{14.5}~{\rm cm}^{-2}$, consistent with our findings for galaxies selected with $D<R_{\rm vir}$. 


\begin{figure*}
    \vskip-0.0cm 
    \centering
    \includegraphics[width=0.33\linewidth]{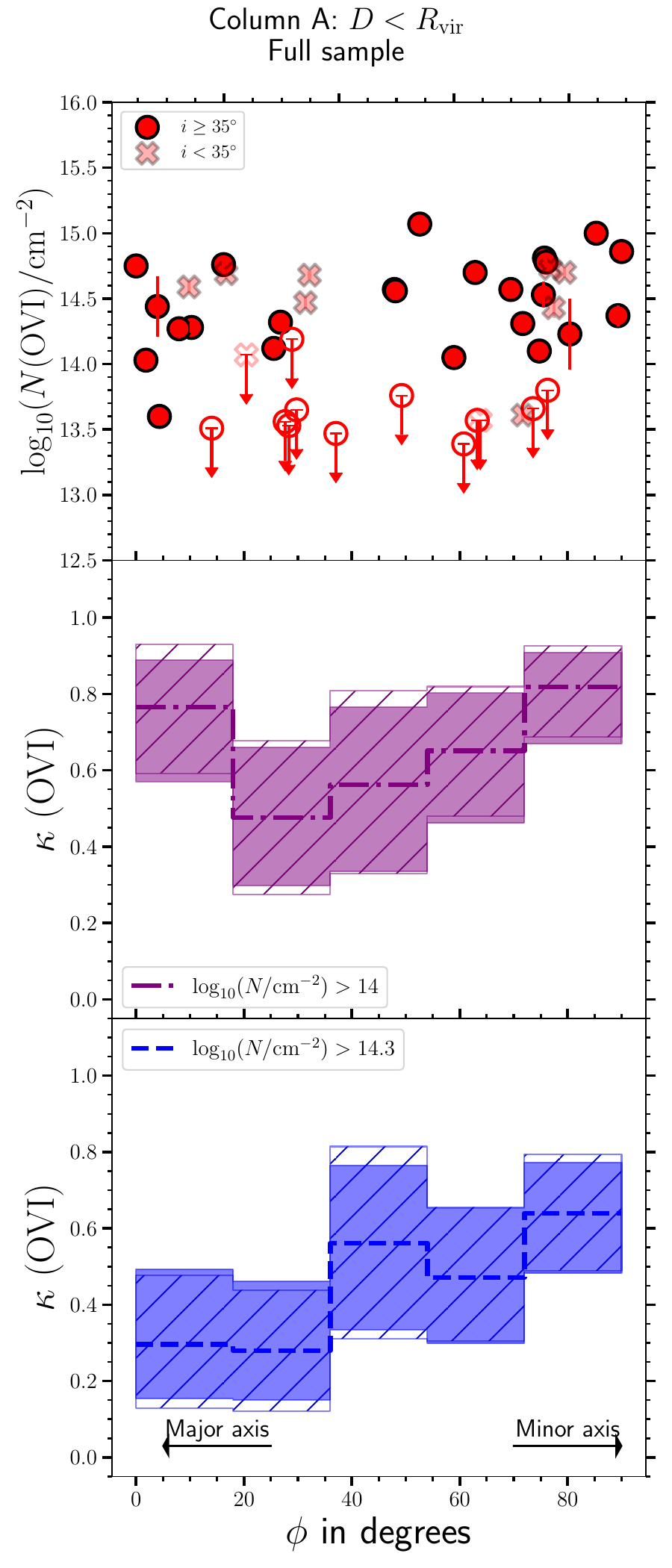}%
    \includegraphics[width=0.33\linewidth]{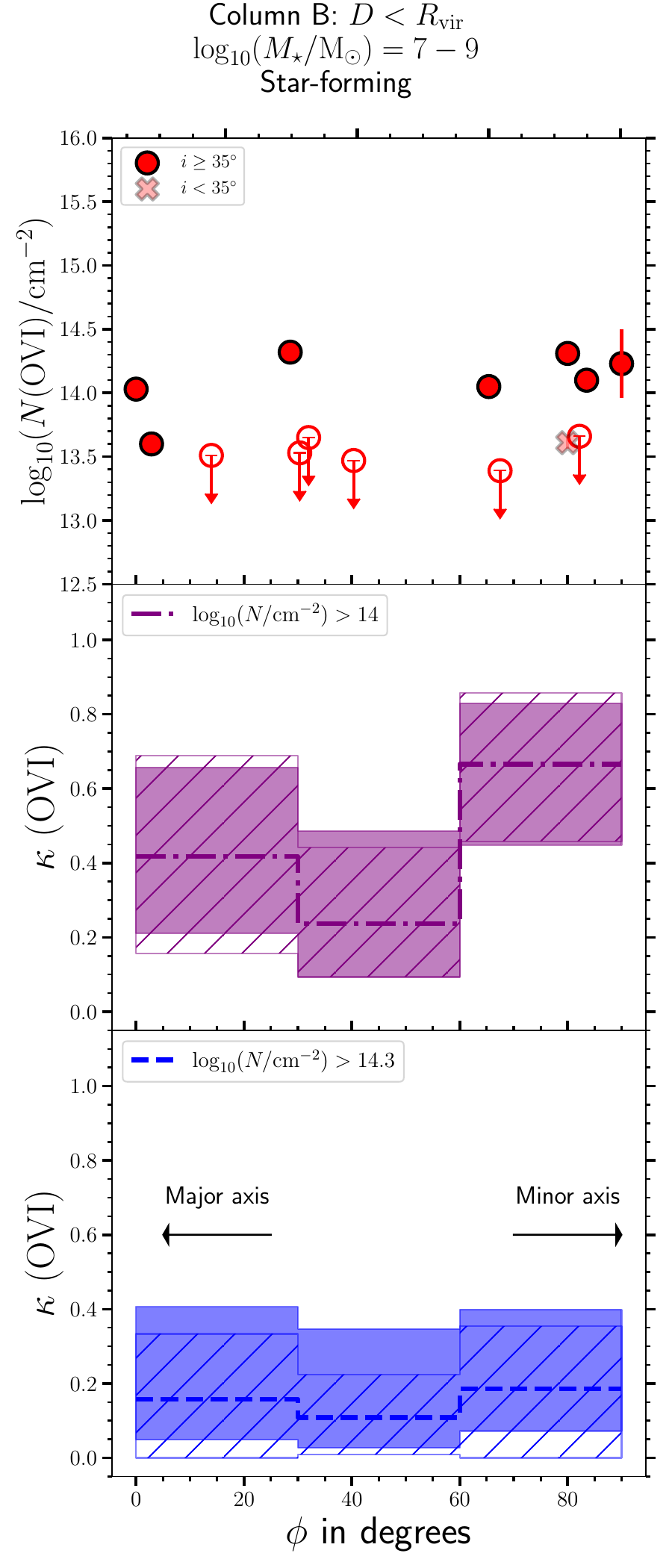}%
    \includegraphics[width=0.33\linewidth]{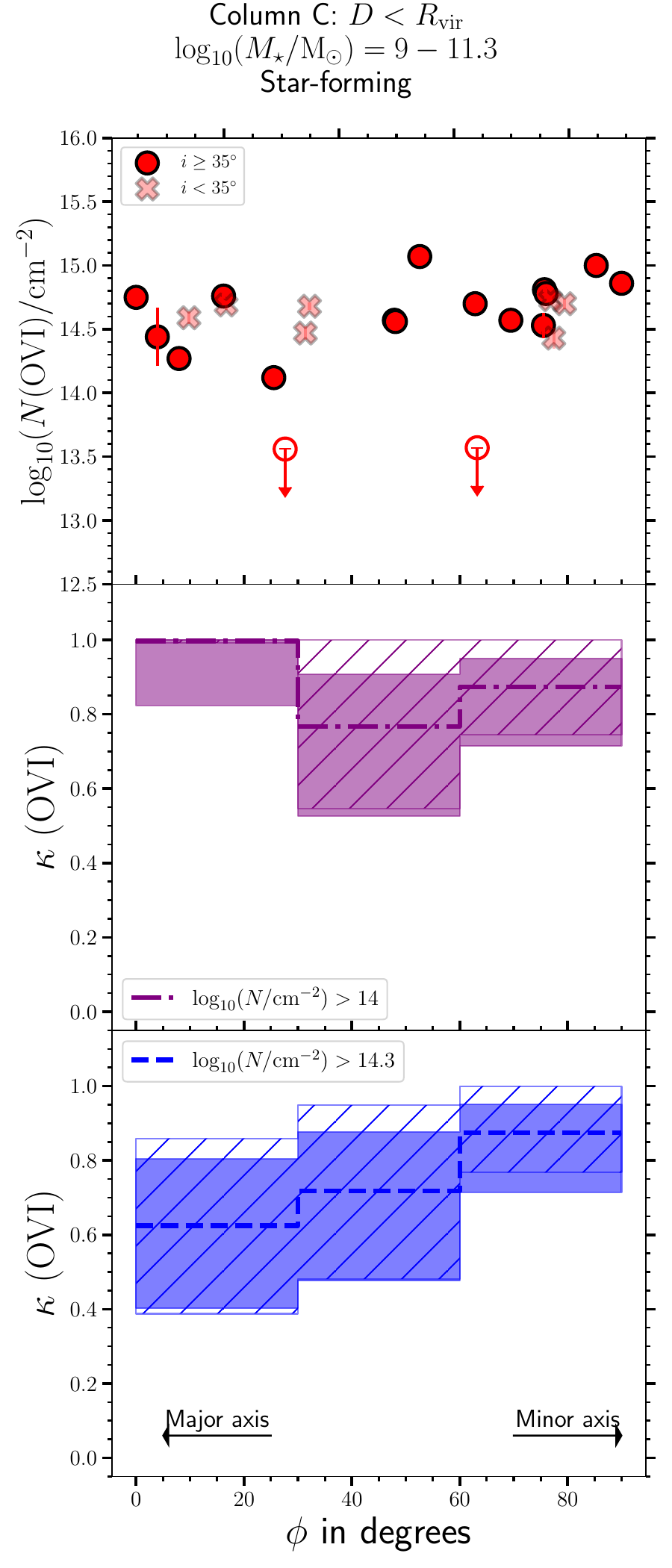}       
    \caption{Same as Fig.~\ref{fig:hi_all} but for \OVI. The column density thresholds are $N(\OVI)=10^{14}~{\rm cm}^{-2}$ and $N(\OVI)=10^{14.3}~{\rm cm}^{-2}$ in the {\tt middle} and {\tt bottom} panels, respectively. In {\tt Columns B} and {\tt C}, only star-forming galaxies are selected. }
    \label{fig:ovi_all}
\end{figure*}

Recently, \citet{Dutta4_2025} showed that the size of the \HI-rich CGM is $\approx 1.5 R_{\rm vir}$ for the MUSEQuBES galaxies spanning the stellar mass range \logm\ $\approx 7$–$10$. Dividing the full sample and dwarf galaxy sample into two bins with $D/R_{\rm vir}$ below and above $1.5$, we find no significant trend between $\kHI$ and $\phi$ for either of the $N(\HI)$ thresholds within or beyond 1.5$R_{\rm vir}$. A similar analysis could not be carried out with the massive subsample due to the lack of massive galaxies with $D>1.5R_{\rm vir}$.

\subsubsection{Azimuthal dependence of \OVI\  in the CGM}  
\label{sec:phi-dep-ovi}

In the top panel of {\tt Column A} in Fig.~\ref{fig:ovi_all}, individual $N(\OVI)$ measurements for the 45 galaxies\footnote{For one galaxy, the \OVI\ doublet falls within the geocoronal \lya\ emission} with $D<R_{\rm vir}$ are plotted against the most probable $\phi$ with solid and hollow red circles, indicating the detections and 3$\sigma$ upper limits for non-detections, respectively. The cross symbols indicate face-on galaxies with most probable $i<35^{\circ}$. 
No significant variation of the detected $N(\OVI)$ with $\phi$ is observed within $R_{\rm vir}$, although the non-detections are primarily at intermediate $\phi$.
In the middle and bottom of {\tt Column A} in Fig.~\ref{fig:ovi_all}, we show the \OVI\ covering fraction (\kOVI) against $\phi$ in 18$^{\circ}$ bins, for the 35 galaxies for threshold $N(\OVI)$ of $10^{14}~{\rm cm}^{-2}$ and $10^{14.3}~{\rm cm}^{-2}$, respectively.
We have discarded 10 face-on galaxies with $i<35^{\circ}$ for the covering fraction analysis.
The \kOVI, for a threshold of $10^{14}~{\rm cm}^{-2}$, is observed to gradually increase towards the major ($\phi\leq18^{\circ}$, $\kOVI=0.76_{-0.19}^{+0.13}$) and minor ($\phi\geq72^{\circ}$, $\kOVI=0.82_{-0.15}^{+0.09}$) axes from the intermediate $\phi$ where the \kOVI\ is the lowest ($\phi=18^{\circ}-36^{\circ}$, $\kOVI=0.47_{-0.17}^{+0.19}$). For the higher threshold of $N(\OVI)=10^{14.3}~{\rm cm}^{-2}$, \kOVI\ exhibit marginal enhancement along the minor axis ($\kOVI=0.64^{+0.13}_{-0.18}$ for $\phi\geq72^{\circ}$ compared to $\kOVI=0.30^{+0.19}_{-0.15}$ for $\phi\leq18^{\circ}$).

Recent works \citep[see e.g.,][]{Tchernyshyov_2022, Dutta2_2025} suggest that for passive galaxies, \kOVI\ is significantly lower than for star-forming galaxies.
Additionally, \OVI\ is argued to arise from different physical processes in low- and high-mass galaxies, which is reflected in both the column density and covering fraction profiles \citep[]{Dutta2_2025}. It is therefore imperative to control for the SFR  and  $M_{\star}$ of the galaxies in which the azimuthal anisotropy of \OVI\ is to be investigated.
We thus chose 38 star-forming galaxies with $D<R_{\rm vir}$ from our sample and divided them into low- and high-mass subsamples with \logm~$=7-9$ and \logm~$=9-11.3$, respectively.

 {\tt Columns B} and {\tt C} of Fig.~\ref{fig:ovi_all} present the same analysis as {\tt Column A}, but separately for the 13 star-forming dwarf galaxies and 17 star-forming high-mass galaxies, respectively. As before, we have discarded one dwarf and 7 massive galaxies with $i<35^{\circ}$ from this analysis.
 The measured $N(\OVI)$ do not show any appreciable variation with $\phi$ for any of the stellar mass bins. However, the \OVI\ non-detections correspond primarily to low-mass galaxies, and intermediate $\phi$. 

For low-mass star-forming galaxies, \kOVI\ exhibits a marginal enhancement at $\phi\lesssim30^{\circ}$ ($\kOVI=0.40^{+0.24}_{-0.20}$) and $\phi\gtrsim60^{\circ}$ ($\kOVI=0.66^{+0.17}_{-0.21}$) compared to intermediate $\phi=30^{\circ}-60^{\circ}$ ($\kOVI=0.24^{+0.26}_{-0.14}$)
for the $N(\OVI)$ threshold of $10^{14}~{\rm cm}^{-2}$.
This marginal enhancement at both low and high $\phi$ for low-mass galaxies suggests a bimodal distribution of \kOVI\ aligned with the major and minor axes. Recently, \citet{Dutta2_2025} argued that the extent of the \OVI-rich CGM is $\approx 0.8R_{\rm vir}$ for dwarf galaxies (\logm~$< 9$). We find that this bimodality of low-mass galaxies persists even when the analysis is restricted to $D/R_{\rm vir} < 0.8$. 
However, we note that the current sample size is relatively modest, and a larger dataset is required to draw more robust and statistically significant conclusions.

The massive, star-forming galaxies with \logm~$=9-11$, on the contrary, almost always exhibit \OVI\ absorption with $N(\OVI)\geq10^{14}~{\rm cm}^{-2}$ within the virial radius irrespective of the azimuthal angle. The \kOVI\ remains roughly uniform across the $\phi$ bins for the threshold $N(\OVI)=10^{14}~{\rm cm}^{-2}$. 
Among the two non-detections, the one at $\phi \approx 65^{\circ}$ is associated with a sub-$L_*$ galaxy with \logm=9.7 at $D/R_{\rm vir} \approx 0.95$. Given the uncertainties in the virial radius estimates, this system may, in fact, lie outside $R_{\rm vir}$, the boundary of \OVI-bearing CGM, as suggested in \citet[]{Dutta2_2025}. 
The other non-detection is associated with a galaxy of \logm$=10.8$, placing it at the high-mass end of the bin. This is consistent with the suppression of $\kappa$ observed in massive, star-forming galaxies with \logm$\gtrsim10.5$, even within $R_{\rm vir}$ \citep[\kOVI\ $\approx50\%$ within $R_{\rm vir}$ for this \logm\ bin, see][]{Dutta2_2025}. These non-detections are therefore more plausibly explained by variations in stellar mass and uncertainties in proximity to the CGM boundary, rather than by a strong dependence on azimuthal angle. As mentioned earlier, systematics due to the inner-CGM measurements for the massive sample aiding $D/R_{\rm vir}\lesssim0.5$ galaxies may influence our conclusion. Splitting the massive, SF galaxies into two bins with $D/R_{\rm vir}=0-0.5$ and $D/R_{\rm vir}=0.5-1$, we do not find any significant azimuthal variation of \kOVI\ in either of the $D/R_{\rm vir}$ bins (Appendix, right panel of Fig.~\ref{fig:app:2dn_bins}). However, we note that the outer-CGM measurements are ill-constrained at $\phi=30^{\circ}-60^{\circ}$ due to the lack of massive, SF samples in this $\phi$ bin. 

A higher $N(\OVI)$ threshold of $10^{14.3}~{\rm cm}^{-2}$ reduces \kOVI\ across all azimuthal angles, though $\kOVI=0.87_{-0.16}^{+0.08}$ is marginally enhanced at $\phi \gtrsim 60^\circ$ compared to $\kOVI=0.62^{+0.18}_{-0.22}$ at $\phi\lesssim30^{\circ}$ ({\tt Column C}; bottom in Fig.~\ref{fig:ovi_all}). This trend persists for the higher threshold of $N(\OVI) = 10^{14.6}~{\rm cm}^{-2}$.

Overall, we find that there is no evidence for anisotropy in the distribution of \OVI\ in the CGM of isolated, star-forming, massive galaxies.


\label{sec:result_nhi}

\subsection{Variation of \OVI\ kinematics with $\phi$} 
\label{sec:result_kin}

\begin{figure}
    \centering
    \includegraphics[width=1\linewidth]{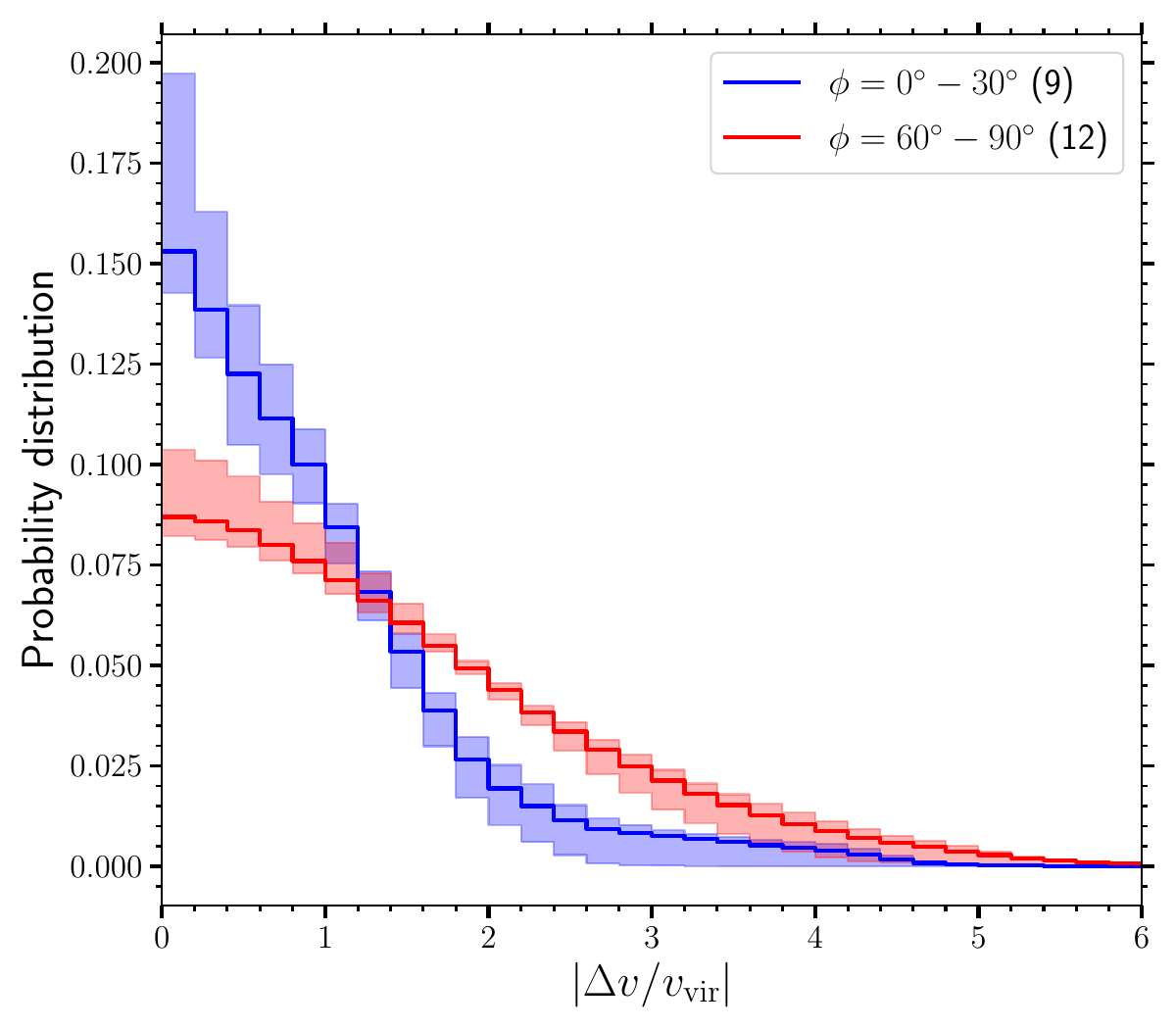}
    \caption{The $v_{\rm vir}-$normalized pixel-velocity two-point correlation functions (TPCFs) for \OVI\ absorbers associated with $D<R_{\rm vir}$ galaxies
    for two bins of azimuthal angle as indicated by the legends. The shaded regions represent the 1$\sigma$ confidence intervals obtained from 200 bootstrap realizations. The numbers in the parentheses indicate the number of galaxies contributing to the plot. 
    Face-on galaxies with $i<35^{\circ}$ are excluded.}  
    \label{fig:ovi_kin}
\end{figure}

To systematically examine the relationship between \OVI\ absorption kinematics and the azimuthal angle of the associated galaxy, we utilize the pixel-velocity two-point correlation function \citep[TPCF;][]{Nielsen_2015,Dutta3_2025}. The pixel-velocity TPCF represents 
a statistical characterization of the velocity dispersion of the absorbing gas.
Following \citet[]{Dutta3_2025}, we first calculate the velocity differences of all
possible pixel pairs using the pixels over which the OVI absorption is detected\footnote{We have used pixels with optical depth $\tau \gtrsim 0.01$.}.
For this analysis, we use the best-fit voigt profiles produced using the best-fit $N(\OVI)$ and $b$, convolved with the COS line spread function (LSF), rather than the observed data. 
Next, for each galaxy, we normalize the pixel velocity differences by the virial velocities of the associated galaxy. This ensures that the kinematic spread in the absorption profile is accounted for by the different halo masses for different galaxies in our sample \citep[see also][]{Dutta3_2025}. The normalized pixel velocities are subsequently binned in $|\Delta v/v_{\rm vir}|$ bins of width 0.2. The resultant histogram is then divided by the total number of pixel pairs to obtain the TPCF.

Fig.~\ref{fig:ovi_kin} shows the normalized \OVI\ TPCF for {\bf two} $\phi$ bins: $\phi=0^{\circ}-30^{\circ}$, and  $60^{\circ}-90^{\circ}$ with blue and red colors, respectively, with 9 and 12 galaxies contributing to each bin.  The intermediate $\phi$ bin of $30^{\circ}-60^{\circ}$ not shown in Fig.~\ref{fig:ovi_kin} contributes only 3 galaxies, and is qualitatively similar to the TPCF of $\phi=60^{\circ}-90^{\circ}$. 
Only galaxies with $D<R_{\rm vir}$ are selected for this analysis. Additionally, face-on galaxies ($i<35^\circ$), for which the azimuthal angle measurements are highly uncertain, are excluded from the analysis. The shaded region indicates the 1$\sigma$ uncertainty obtained from 200 bootstrap realizations. 
 The TPCF in the two bins shows a narrower profile at lower $\Phi$.  
We measure a $\approx6\sigma$ difference between the TPCF of the $\phi=0^{\circ}$–$30^{\circ}$ and $\phi=60^{\circ}$–$90^{\circ}$ bins.

We do not find any statistically significant difference in the stellar mass distribution of galaxies at $\phi<30^{\circ}$ and $\phi>60^{\circ}$ (a 2-sample KS test results in $p\approx0.9$). 
\citet{Dutta3_2025} reported that the virial-velocity-normalized \OVI\ TPCF does not exhibit any significant stellar mass dependence. Therefore, the $\approx6\sigma$ narrowing of the TPCF for absorbers along the galaxy's major axis is unlikely to be driven by differences in the stellar mass distribution of the host galaxies.
 However, the median sSFR (SFR) of $10^{-9.7}~{\rm yr}^{-1}$ ($10^{-0.4}$~\Msun$~{\rm yr}^{-1}$) is marginally higher for galaxies at $\phi\geq 60^{\circ}$ compared to $10^{-10}~{\rm yr}^{-1}$ ($10^{-0.5}$~\Msun${\rm yr}^{-1}$) obtained for galaxies at $\phi< 30^{\circ}$. The $p-$value for the 2-sample KS test reveals a marginal ($p\approx0.05$) difference in sSFR distribution for galaxies at $\phi\leq30^{\circ}$ and $\phi\geq60^{\circ}$. However, the difference is not significant for the SFR distribution ($p\approx0.1$).


Finally, we note that performing the same analysis with \HI\ does not reveal any significant difference in absorber kinematics between the minor and major axes. The lack of difference in \HI\ absorption kinematics is partly due to the saturation of \lya\ (and \lyb) absorbers at moderate $N(\HI)\sim10^{14}~{\rm cm}^{-2}$ ($N(\HI)\sim10^{15}~{\rm cm}^{-2}$). Hence, the kinematic width traced by TPCFs for \HI\ is mostly driven by column density.

\section{Discussion} 
\label{sec:disc}

\subsection{Azimuthal dependence of cool gas in the CGM}

In {\tt Column A} of Fig.~\ref{fig:hi_all}, a marginal suppression of \kHI\ for the threshold $N(\HI)=10^{14.5}~{\rm cm}^{-2}$ within $R_{\rm vir}$ is observed at intermediate $\phi$ for the full sample, primarily driven by two non-detections at $\phi \approx 30^{\circ} - 40^{\circ}$. These non-detections are associated with low-mass, dwarf galaxies (top panel of {\tt Column B}), leading to a significant suppression of \kHI\ for the threshold $N(\HI)=10^{14.5}~{\rm cm}^{-2}$ in the intermediate-$\phi$ bin for low-mass galaxies (middle panel of {\tt Column B}). In contrast, edge-on ($i > 35^{\circ}$), high-mass galaxies show \HI\ absorption at all $\phi$ (top panel of {\tt Column C}), resulting in a uniform, near-unity \kHI\ within $R_{\rm vir}$ for this threshold with no significant azimuthal variation.
For the higher threshold of $N(\HI) = 10^{15}~{\rm cm}^{-2}$, no significant azimuthal variation in \kHI\ is observed for either stellar mass bin and for the full sample.

\citet{Borthakur_2015} found no correlation between azimuthal angle and the \lya\ rest-frame equivalent width. Similarly, 
\citet{Pointon_2019} reported no appreciable azimuthal dependence of the circumgalactic \HI\ column density. Both of these studies probe galaxies over an impact parameter range similar to that of this work ($\lesssim 200$~pkpc) with similar sample size (45 for \citet[]{Borthakur_2015} and 47 for \citet[]{Pointon_2019}) but focus primarily on high-mass systems ($\approx 10^{10}~$\Msun). The high mass galaxies in our sample show a qualitatively similar behavior as evident from {\tt Column C} of Fig.~\ref{fig:hi_all}. 
 \citet[]{Beckett_2021} reported a bimodality in \HI\ absorption for their sample of $\approx300$ quasar-galaxy pairs in the Q0107 triplet sightline with \logm$\approx8-12$, the significance of which increases when restricting to the `strong' (${\rm log}_{10}(N(\HI)/{\rm cm}^{-2}) >14$) \HI\ absorbers. 
However, unlike determining the covering fraction as a function of $\phi$, they used a Hartisan's dip test on the $\phi$ distribution of galaxies with detected \HI\ absorber to report the bimodality. We have refrained from carrying out a similar analysis, as Hartigan’s dip test suggests a departure from unimodality ($p\sim0.05$) for the overall $\phi$ distribution (see Fig.~\ref{fig:galprop_morph})\footnote{However, the KS and Anderson–Darling tests do not reject consistency with a uniform distribution ($p\sim0.2$).}.
However, a significant suppression/dip in \kHI\ is observed at intermediate $\phi$ in our sample, but only for low-mass galaxies within $R_{\rm vir}$ for a threshold $N(\HI)=10^{14.5}~{\rm cm}^{-2}$.

Low-ionization metal lines, such as \MgII, are well-known tracers of cool gas in the CGM. \citet[]{Guo_2023} showed the presence of anisotropic \MgII-bearing CGM in emission around massive (\logm\ $>9.5$) galaxies by stacking $\approx112$ edge-on galaxies. They reported a lack of anisotropy for the low-mass (\logm\ $<9.5$) galaxies. However, this emission-based measurements are confined to $\sim 10$ pkpc, as opposed to the absorption-line measurements at $D\gtrsim50$ pkpc presented in this work.
 Based on a sample of 10 $\sim L_*$ galaxies at $z\approx 0.1$ exhibiting \MgII\ absorption with $W_r>0.3~$\AA\ in the background quasar spectra, \citet[]{Bouche_2012} showed that the azimuthal distribution of quasar sightlines is bimodal, with a preference for alignment along the projected major ($\phi \lesssim 20^{\circ}$) and minor ($\phi \gtrsim 60^{\circ}$) axes of galaxies. \citet[]{Kacprzak_2012} reached a similar conclusion with a larger sample of 88 \MgII-absorption selected ($W_r>0.1~$\AA) galaxies with $z\approx0.1-1.1$. Additionally, they found that the galaxies without \MgII\ absorption do not exhibit any preferential $\phi$ values.  
On the contrary, stacking $\gtrsim5000$ background galaxy spectra, \citet[]{Bordoloi_2011} showed an enhanced stacked \MgII\ equivalent width along the minor axis at impact parameters $\lesssim 50$ pkpc compared to the major axis. However, no such enhancement was observed at larger radii.

Based on a galaxy-centric study of \MgII\ absorbers in the CGM of isolated galaxies, \citet[]{Cherrey_2025} found that the quasar sightlines for the \MgII\ absorbing galaxies are preferentially located along the minor axis, while the sightlines for the non-absorbing galaxies are preferentially observed along the intermediate azimuthal angles $(30^{\circ} < \phi < 60^{\circ})$. On the contrary, \citet[]{Huang_2021} reported a lack of azimuthal angle dependence of the covering fraction and $W_r$ of \MgII\ absorbers based on a sample of isolated and star-forming galaxies with $D<0.4R_{\rm vir}$. 
We note that the galaxies in \citet[]{Huang_2021} are primarily massive ($\gtrsim10^{9}$ \Msun\ with median $\approx10^{10}$ \Msun).
The lack of azimuthal dependence of the cool gas for the massive galaxies in our sample is consistent with the findings of \citet[]{Huang_2021}, although their work focuses on galaxies with $D/R_{\rm vir}<0.4$. It is worthwhile to note that the lack of azimuthal variation of  $W_r$ (\MgII) and $\phi$ is also reported in \citet[]{Dutta_2020} for galaxies that are primarily in pairs or groups.

The lack of consensus regarding azimuthal variations in \MgII\ absorption may stem from differences in survey strategies. While studies such as \citet[]{Bouche_2012}, \citet[]{Kacprzak_2012} adopt an absorption-centric approach, others like \citet[]{Dutta_2020}, \citet[]{Huang_2021}, \citet[]{Cherrey_2025} are galaxy-centric CGM surveys. Moreover, additional galaxy properties - such as SFR, $M_{\star}$, impact parameter, and environment significantly influence the \MgII\ covering fraction \citep[e.g.,][]{Dutta_2020, Dutta_2021, Cherrey_2024, Cherrey_2025}, potentially modulating its azimuthal dependence. Therefore, controlling for these galaxy parameters is essential when investigating azimuthal trends. 

Finally, we emphasize that it is not straightforward to compare our \HI\ results with the seemingly contradictory findings from \MgII\ studies. Although both ions trace cool, photoionized gas in the CGM, the relationship between \MgII\ and \HI\ column densities is not well defined - particularly in the low column density regime. For instance, \citet[]{Lan_2017} reported an empirical correlation between $W_r$(\MgII) and $N(\HI)$ for optically thick absorbers albeit with a significant scatter, spanning several orders of magnitude in $N(\HI)$ for a given $W_r$(\MgII). In this context, we point out that we do not find any azimuthal angle dependence in \kHI\ for optically thick gas, using a threshold of $N(\HI) > 10^{17} {\rm cm}^{-2}$.


\subsection{Azimuthal dependence of highly-ionized, warm-hot gas in the CGM} 

The Column {\tt A} of Fig.~\ref{fig:ovi_all} showed a gradual decline of \kOVI\ for threshold $N(\OVI)=10^{14}~{\rm cm}^{-2}$ toward the intermediate $\phi$ bins from the higher- and lower-$\phi$ bins for the full sample within $D<R_{\rm vir}$. Upon restricting the sample to include only star-forming galaxies and dividing it into low- and high-mass subsamples, we found that a similar, tentative trend is observed for the star-forming, low-mass (\logm$\lesssim9$) galaxies ({\tt Column B}, Fig.~\ref{fig:ovi_all}). However, no significant azimuthal variation of \kOVI\ with this threshold is observed for the massive, star-forming subsample ({\tt Column C}, Fig.~\ref{fig:ovi_all}). As discussed in sect. \ref{sec:phi-dep-ovi}, the two non-detections in this stellar mass bin can be attributed to the role of stellar mass and uncertain CGM boundary proximity, and not necessarily to the azimuthal angle. 

For the higher threshold of $N(\OVI)=10^{14.3}~{\rm cm}^{-2}$,  \kOVI\ for the full sample increases with $\phi$. However, as the low-mass galaxies do not exhibit $N(\OVI)$ above this threshold, this trend is primarily driven by the massive, star-forming subsample as evident from {\tt Column C} of Fig.~\ref{fig:ovi_all}. 
Although the detected $N(\OVI)$ does not show any significant correlation with $\phi$ (Kendall-$\tau$ test reveals $p\approx0.09$), the mean $N(\OVI)=10^{14.5}~{\rm cm}^{-2}$ at $\phi<45^{\circ}$ is marginally lower than the mean $N(\OVI)=10^{14.8}~{\rm cm}^{-2}$ at $\phi\geq 45^{\circ}$.

Previously, \citet[]{Kacprzak_2015} reported an azimuthal dependence of \OVI\ absorption around galaxies, wherein \kOVI\ (for a threshold rest-frame equivalent width of 0.1~\AA, corresponding to $N(\OVI)\approx10^{13.9}~{\rm cm}^{-2}$) was significantly enhanced along the major ($\phi\approx10^{\circ}-20^{\circ}$) and minor ($\phi\gtrsim60^{\circ}$) axes. 
For a similar threshold of $N(\OVI)=10^{14}~{\rm cm}^{-2}$, we do not observe any enhancement of \kOVI\ along the major or minor axes for the massive, star-forming galaxies with $D<R_{\rm vir}$ in our sample. Across all azimuthal angles, we find $\kappa_{\mathrm{OVI}}$ to be consistently high (80–100\%), indicating an isotropic distribution of \OVI-bearing gas.
We emphasize that \citet[]{Kacprzak_2015} did not control for the stellar mass, star-formation rates, and impact parameter, all of which can significantly influence the \OVI\ covering fraction. Indeed, the \kOVI$\approx30\%$ at $\phi\approx20^{\circ}-60^{\circ}$ reported in \citet[]{Kacprzak_2015} is significantly lower than the \kOVI$\approx 80\%-100\%$ for the massive galaxies within $D<R_{\rm vir}$ at similar azimuthal angles that we find for a similar threshold.

\citet{Dutta2_2025} argued that the origin of \OVI\ in the CGM of dwarf galaxies is likely different than for massive galaxies.
A collisionally ionized ambient halo, with conservative limits on the CGM baryon fraction ($\approx15\%$) and metallicity (1/3 solar), can account for the observed mass of the \OVI-bearing CGM of massive galaxies. The near-unity \OVI\ covering fraction across all $\phi$ values is indicative of such a volume-filling, warm-hot phase for the massive, star-forming galaxies.
 We note that the near-unity covering fraction within the $R_{\rm vir}$ for the massive SF galaxies does not necessarily imply a near-unity volume-filling fraction. While mixing layers between $>10^6$ K virialized gas and $\sim 10^4$K cool gas are often argued as the origin of circumgalactic \OVI, we note that the median stellar mass of the massive SF sample (\logm$\sim10^{10}$\Msun) in this work is too low to produce virial temperature of $>10^6$ K. Additionally, due to the nonthermal pressure support (e.g., turbulence, cosmic rays), the
virial temperature for a given halo mass can be overestimated \citep[][]{Lochhaas_2021, Tchernyshyov_2023}. On the contrary, the virial temperature of $\sim 10^{5.5}$ K for these halos is ideal to produce \OVI\ in CIE.
Cosmological hydrodynamical simulations have predicted a volume-filling, CIE origin of circumgalactic \OVI\ \citep[EAGLE,][]{Oppenheimer_2016} for a similar halo-mass range. \citet{Lehner_2020} further showed that, in comparison to these simulations, their observations of the CGM of M31 can be explained by \OVI\ produced in a volume-filling, virialized phase.
 The near-unity \OVI\ covering fraction across all $\phi$ values, hence, is indicative of such a volume-filling, warm-hot phase for the massive, star-forming galaxies. 
On the contrary, non-equilibrium processes, galactic outflows, and photoionization are thought to play a crucial role in producing \OVI\ in the CGM of star-forming dwarf galaxies. The observed marginal azimuthal dependence of the \OVI\ covering fraction is consistent with this picture, with the enhancements along the minor and major axes likely arising from warm, metal-enriched outflows and inflows, respectively.

\subsection{Azimuthal dependence of \OVI\ kinematics in the CGM}  

In Fig.~\ref{fig:ovi_kin}, we showed that the \OVI\  virial-velocity-normalized TPCF for the galaxies along the minor axis is significantly ($\gtrsim6\sigma$) wider compared to that along the major axis, indicating larger velocity dispersions.

The enhanced velocity dispersion along the minor axis could be an indicator of gas entrained in outflows with large kinematic spreads. \OVI\ systems with large velocity spreads have been argued to be associated with galactic-scale outflows \citep[see e.g.,][]{Tripp_2011, Muzahid_2014, Muzahid_2015}.  Down-the-barrel \OVI\ absorption in local starburst galaxies \citep[see][]{Grimes_2009} also tends to show significantly broader absorption profiles compared to \OVI\ absorbers detected in `blind' absorption line surveys \citep[]{Tripp_2008, Muzahid_2012, Danforth_2016}.
 Conversely, the accreting or rotating gas around galaxies may be more kinematically quiescent, with velocity dispersion scaling with the circular velocity. The kinematics of the \OVI-bearing gas phase thus appear to be sensitive to galaxy orientation, suggesting the presence of anisotropic gas flows around galaxies.


\begin{figure}
    \centering
    \includegraphics[width=1\linewidth]{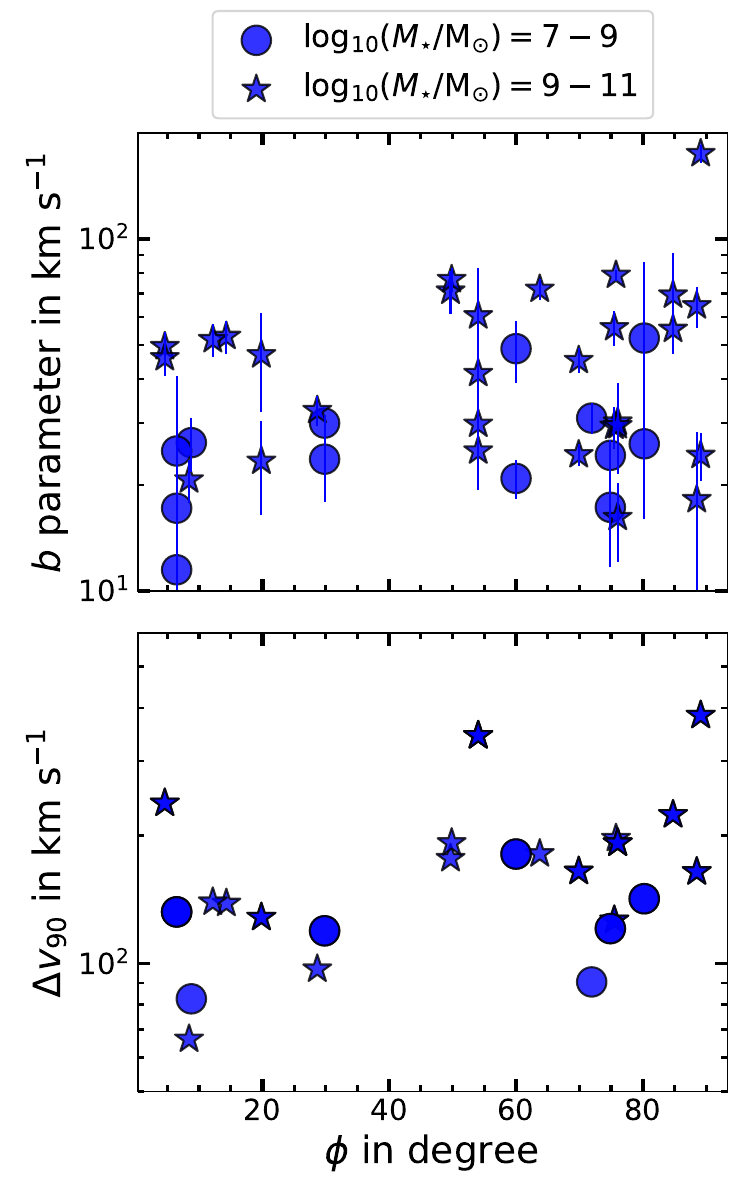}
    \caption{Doppler-$b$ parameters for the \OVI\ components ({\tt top}) and $\Delta v_{90}$ of the {\it systems} ({\tt bottom}) associated with the galaxies with $i>35^{\circ}$ and $D<R_{\rm vir}$ plotted against $\phi$. In both panels, the circles and star symbols indicate the components associated with low-mass and high-mass galaxies, respectively. }
    \label{fig:b_phi_ovi}
\end{figure}

 Although our result seems to be in conflict with \citet[]{Nielsen_2017}, who reported a kinematically uniform \OVI\ absorption around their galaxy sample, the reported kinematic uniformity was only robust for their full sample without any inclination angle cut. A wider \OVI\ TPCF along the minor axis was previously observed by \citet[]{Nielsen_2017} for their edge-on ($i\geq 51^{\circ}$) galaxy sample, with a $4.6\sigma$ significance, which reduces to $\approx3\sigma$ when a strong outlier absorber along the minor axis is discarded. 
We have verified that the normalized TPCF in the two $\phi$ bins does not show appreciable difference once we relax the inclination angle cut of $i>35^{\circ}$. The lack of halo mass dependence in the normalized TPCF for \OVI\ absorbers for edge-on galaxies has been previously reported by \citet[][]{Ng_2019}. However, we have verified that the halo mass distribution in the two azimuthal angle bins is not significantly different.

Besides the TPCF, the $b$ parameter of the individual component and $\Delta v_{90}$, defined as the extent within which 90\% of the total column density of a {\it system} is contained, are often used as proxies for the kinematic width of the absorption.  
 The top and bottom panels of Fig.~\ref{fig:b_phi_ovi} show the Doppler-$b$ parameters and $\Delta v_{90}$ plotted against $\phi$ for the \OVI\ absorbers associated with galaxies with $i>35^{\circ}$ and $D<R_{\rm vir}$. The circles and star symbols indicate the components associated with low-mass and high-mass galaxies, respectively. While the $b$ parameter distributions for galaxies with $\phi<30^{\circ}$ and $\phi>60^{\circ}$ do not
 differ significantly ($p\approx0.3$), the $\Delta v_{90}$ in these two bins exhibit marginally significant difference in a 2-sample KS test ($p\approx0.02$), wherein the median [68\%  interval] of $\Delta v_{90}$ are 128 \kms\ [86-139 \kms] and 173 \kms\ [125-203 \kms] along the disk and pole, respectively, reflecting a broader kinematic profile of \OVI\ absorbers along the polar region. 

 In summary, our results show that the covering fraction of \OVI\ is uniformly high across all azimuthal angles for the high-mass galaxy subsample, indicating an isotropic distribution of highly ionized, metal-enriched gas in the halo. This suggests that the origin of \OVI\ is primarily governed by global halo conditions such as the virial temperature. However, the kinematics of \OVI\ reveal significantly higher velocity dispersion along the minor axis compared to the major axis. This implies that while \OVI\ is omnipresent throughout the halo of star-forming, \logm$\approx 9-11$, its kinematic properties retain information about where in the halo it was produced. The observed higher velocity dispersion of  \OVI\ along the minor axis is likely related to galactic-scale outflows that preferentially emerge perpendicular to the disk plane. Thus, \OVI\ traces both the widespread ionized CGM and the underlying dynamical complexity associated with galaxy-scale gas flows. 


 \section{Summary} 
 \label{sec:summary}

 In this paper, we have presented a detailed investigation on the relationship between galaxy orientation and circumgalactic \HI\ and \OVI\ absorption, utilizing high-resolution $HST$ imaging of 113 galaxies, including 91 primarily low-mass galaxies from the MUSEQuBES \citep[]{Dutta1_2024} and 22 high-mass galaxies from the COS-Halos \citep[]{Tumlinson_2013} surveys. The galaxies span a redshift range (68\%) of $z=0.2-0.5$, stellar mass range of \logm\ = 7.9-10.1, star-formation rate of $10^{-1.6}-10^{0.3}~M_{\star}~{\rm yr}^{-1}$ with impact parameters $<$300~pkpc. The galaxies are selected to be isolated with no companion galaxies within 500 pkpc and 500 \kms.  We present a novel {\tt Python} wrapper for the widely used {\tt GALFIT} software, designed to systematically explore the input parameter space using a Bayesian approach and to determine realistic uncertainties on the best-fit model parameters. 
By incorporating the posterior distributions of galaxy azimuthal angles, we examined the trends between column density and covering fraction of \HI\ and \OVI\ absorption in the CGM as a function of azimuthal angle. We adopt the convention that azimuthal angle $\phi$ of 0$^{\circ}$ and 90$^{\circ}$ correspond to the projected major and minor axes of the galaxies, respectively. Our main findings are- 

 \begin{itemize}
     \item  We find that the \HI\ covering fraction, \kHI, within $R_{\rm vir}$ for the full galaxy sample exhibits marginal suppression at intermediate azimuthal angle $\phi$ of $\approx20^{\circ}-70^{\circ}$ for the threshold $N(\HI)=10^{14.5}~{\rm cm}^{-2}$ ({\tt Column A} of Fig.~\ref{fig:hi_all}).     
     Dividing our galaxy sample into low- and high-mass bins, we find that this is primarily driven by two non-detections at $\phi\approx30^{\circ}-40^{\circ}$ in the low-mass subsample, leading to a significant suppression of \kHI\ for the same threshold in the intermediate $\phi$ bin ({\tt Column B} of Fig.~\ref{fig:hi_all}). A higher threshold of $N(\HI)=10^{15}~{\rm cm}^{-2}$ does not reveal any significant $\phi-$dependence of \kHI\ for the full sample or the low-mass subsample. 
     No significant azimuthal variation of \kHI\ is observed for the high-mass subsample ({\tt Column C} Fig.~\ref{fig:hi_all}) for either of the adopted thresholds. We do not observe any significant azimuthal dependence beyond the virial radius of the galaxies.

     \item The \kOVI\ with threshold $N(\OVI)=10^{14}~{\rm cm}^{-2}$ within $R_{\rm vir}$ for the full galaxy sample exhibits a gradual decline toward the intermediate $\phi$ bins from the high- and low-$\phi$ bins (Fig.~\ref{fig:ovi_all}, {\tt Column A}). A similar, tentative trend is observed for the star-forming, low-mass (\logm$\lesssim9$) galaxies ({\tt Column B} in Fig.~\ref{fig:ovi_all}). \kOVI\ of the high-mass, star-forming galaxies within the virial radius does not exhibit any significant trend with $\phi$ for this threshold ({\tt Column C} in Fig.~\ref{fig:ovi_all}). For a higher threshold of $N(\OVI)=10^{14.3}~{\rm cm}^{-2}$, the \kOVI\ for the full sample exhibits a gradual increase with increasing $\phi$, which is primarily driven by the massive star-forming subsample.

     \item 
     We use the pixel-velocity two-point correlation function (TPCF) normalized by the circular velocity of the host halo to investigate the relationship between \OVI\ kinematics and azimuthal angle.
     We find that the normalized TPCF exhibit $\approx 6\sigma$ narrowing for the absorbers along the major axis ($\phi\lesssim 30^{\circ}$) compared to absorbers at $\phi=60^{\circ}-90^{\circ}$ (Fig.~\ref{fig:ovi_kin}). This is consistent with the overall higher $\Delta v_{90}$ of \OVI\ absorbers along the polar ($\phi=60^{\circ}-90^{\circ}$) direction (Fig.~\ref{fig:b_phi_ovi}).

 \end{itemize}

The azimuthal dependence of the CGM, as inferred from quasar absorption-line studies, remains a subject of considerable debate in the literature. Growing observational evidence suggests that galaxy properties such as SFR, $M_{\star}$, $D$, and environment significantly influence the CGM,  rendering azimuthal variations a higher-order effect that requires careful control of these parameters. In this work, we revisit the azimuthal dependence of the \HI\ and \OVI\ absorption while controlling for $M_{\star}$, $D$, SFR, and environment. 
We find no significant azimuthal dependence in the distribution of \HI\ and \OVI\ in the CGM of high-mass galaxies. However, the kinematics of \OVI\ absorption are significantly broader along the projected minor axis. These observations suggest that \OVI\ arises from a volume-filling gas phase governed by global halo conditions, while its kinematics retain information about the site of origin. 
We find tentative evidence that both the cool, neutral gas traced by \HI\ and the highly ionized gas traced by \OVI\ exhibit enhanced covering fractions along the projected major ($\phi<30^{\circ}$) and minor axes ($\phi>60^{\circ}$), relative to intermediate azimuthal angles but only in low-mass dwarf galaxies. A larger sample of dwarf galaxies is required to draw statistically robust conclusions.




\begin{acknowledgments}
We thank Marijke Segers, Lorrie Straka, and Monica Turner for their early contributions to the MUSEQuBES project. SD and SM thank R. Srianand and Aseem Paranjape for insightful discussions. RA acknowledges funding from the European Research Council (ERC) under the European Union's Horizon 2020 research and innovation programme (grant agreement 101020943, SPECMAP-CGM). 

\end{acknowledgments}

\appendix

\section{Summary of the $HST$ observations used in this work} 
\label{sec:hst-summary}

The Table \ref{tab:hst_obs} summarizes the details of $HST$ observations of the quasar fields used in this study. 

\begin{table*}
\centering
\caption{Summary of the HST observation of the quasar fields used in this study}   
\label{tab:hst_obs}
\begin{tabular}{lrrrrr}
 \hline
 Quasar Name & Detector & Filter & Survey & PID & Exp Time (s)\\
 \hline
 \hline
3C57 & ACS/WFC1 & F814W & MUSEQuBES & 14660 & 2179 \\  
FIRST-J020930.7-043826 & ACS/WFC1 & F814W & MUSEQuBES & 14660 & 2171 \\  
HB89-0107-025-NED05 & ACS/WFC1 & F814W & MUSEQuBES & 14660 & 2171  \\  
HB89-0232-042 & ACS/WFC1 & F814W & MUSEQuBES & 14269 & 2180 \\  
HE0153-4520 & ACS/WFC1 & F814W & MUSEQuBES & 13024 & 1200 \\  
HE0226-4110 & ACS/WFC1 & F814W & MUSEQuBES & 13024 & 1200\\  
HE0238-1904 & ACS/WFC1 & F814W & MUSEQuBES & 14660 & 2182 \\  
HE0439-5254 & ACS/WFC1 & F814W & MUSEQuBES & 14269 & 2376 \\ 
LBQS-0107-0235 & ACS/WFC1 & F814W & MUSEQuBES & 14660 & 2171 \\ 
LBQS-1435-0134 & WFC3 & F140W & MUSEQuBES & 14594 & 555 \\ 
PG-1522+101 & ACS/WFC1 & F814W & MUSEQuBES & 14269 & 2184 \\ 
PKS0405-123 & ACS/WFC1 & F814W & MUSEQuBES & 13024 & 1200 \\ 
PKS0552-640 & ACS/WFC1 & F814W & MUSEQuBES & 14269 & 2448 \\ 
QSO-J1009+0713 & WFC3 & F390W & COS-Halos & 11598 & 2370 \\ 
SDSS-J100535.24+013445.7 & ACS/WFC1 & F814W & MUSEQuBES & 14269 & 2180 \\ 
SDSS-J135726.27+043541.4 & ACS/WFC1 & F814W & MUSEQuBES & 14660 & 2171\\ 
SDSSJ091440.38+282330.6 & ACS/WFC1 & F814W & COS-Halos & 13024 & 1200\\ 
SDSSJ092837.98+602521.0 & WFC3 & F160W & COS-Halos & 15975 & 2385 \\ 
SDSSJ094331.61+053131.4 & ACS/WFC1 & F814W & COS-Halos & 13024 & 1200 \\ 
SDSSJ095000.73+483129.3 & ACS/WFC1 & F814W & COS-Halos & 13024 & 1200\\ 
SDSSJ101622.60+470643.3 & WFC3 & F390W & COS-Halos & 16742 & 2762 \\ 
SDSSJ111239.11+353928.2 & WFC3 & F390W & COS-Halos & 16742 & 2676 \\ 
SDSSJ113327.78+032719.1 & ACS/WFC1 & F814W & COS-Halos & 13024 & 1200\\ 
SDSSJ123304.05-003134.1 & ACS/WFC1 & F814W & COS-Halos & 13024 & 1200\\ 
SDSSJ124154.02+572107.3 & ACS/WFC1 & F814W & COS-Halos & 13024 & 1200\\ 
SDSSJ132222.68+464535.2 & ACS/WFC1 & F814W & COS-Halos & 13024 & 1200 \\ 
SDSSJ141910.20+420746.9 & ACS/WFC1 & F814W & COS-Halos & 14269 & 2280\\ 
SDSSJ143511.53+360437.2 & ACS/WFC1 & F814W & COS-Halos & 14269 & 2240\\ 
SDSSJ155504.39+362848.0 & ACS/WFC1 & F814W & COS-Halos & 13024 & 1200 \\ 
SDSSJ161916.54+334238.4 & ACS/WFC1 & F814W & COS-Halos & 14269 & 2216\\ 

 \hline
 \hline
\end{tabular}
\end{table*}


\section{Discarded galaxy sample}  
\label{sec:disc-sample}

\begin{figure}
    \centering
    \includegraphics[width=1\linewidth]{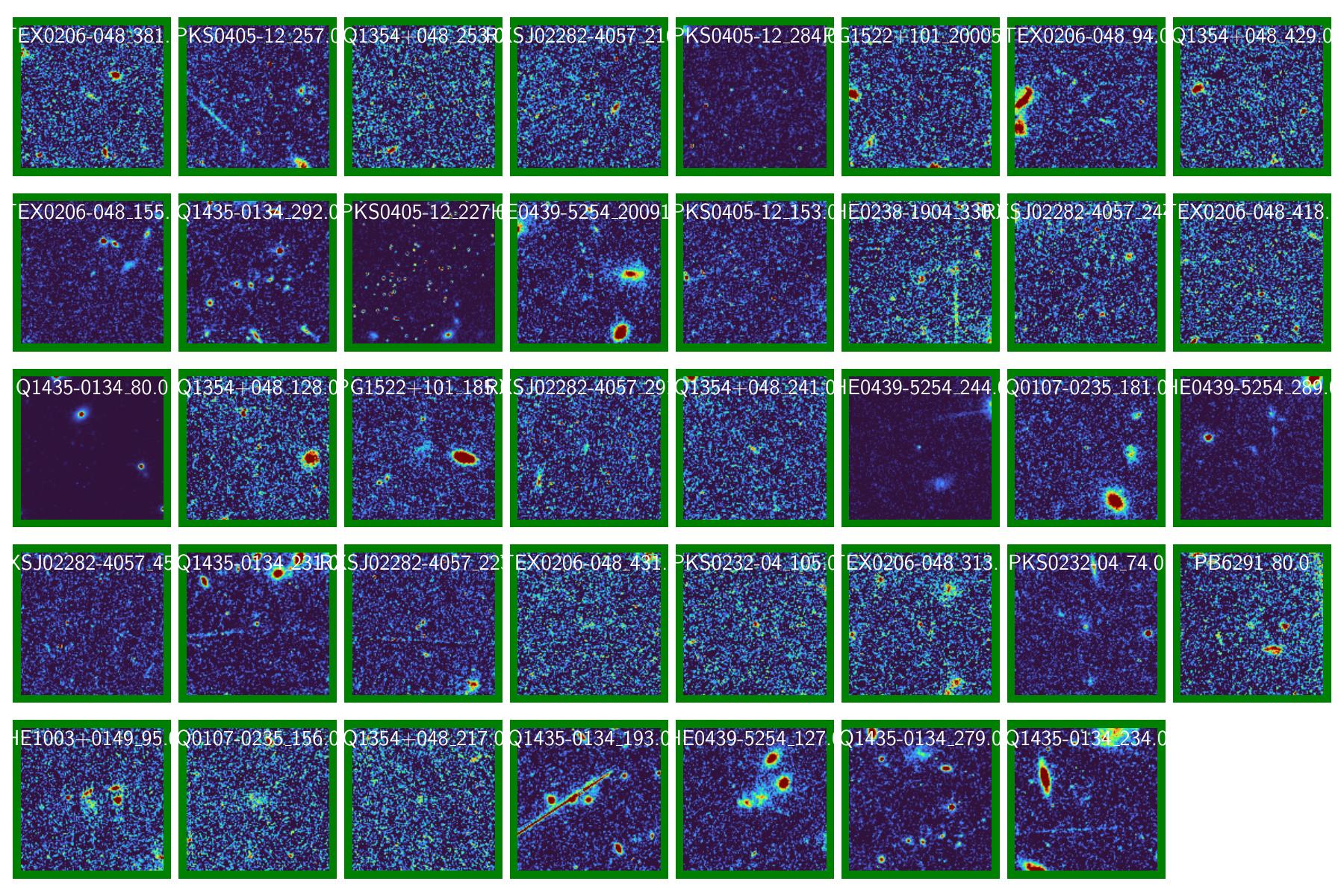}
    \caption{Cutouts of the 39 discarded galaxies from the analysis. The galaxies primarily have lower $S/N$ ($\lesssim10$) than the galaxies selected in this work. The 8 objects with $S/N\gtrsim10$ are either unresolved (most probable $R_e\sim 1$ pixel, e.g., HE0439-5254-244.0), or suffer from projected contamination (e.g., HE0439-5254-127). }
    \label{fig:F_disc}
\end{figure}

In section \ref{sec:gal-sample}, we mentioned that 39 MUSEQuBES galaxies are discarded from our analysis as they are either not detected or unresolved in the $HST$ images. The 201$\times$201 pixel cutouts of the 39 galaxies are shown in Fig.~\ref{fig:F_disc}. In this section, we discuss the detection significance of the galaxies. 

\begin{figure}
    \centering
    \includegraphics[width=0.5\linewidth]{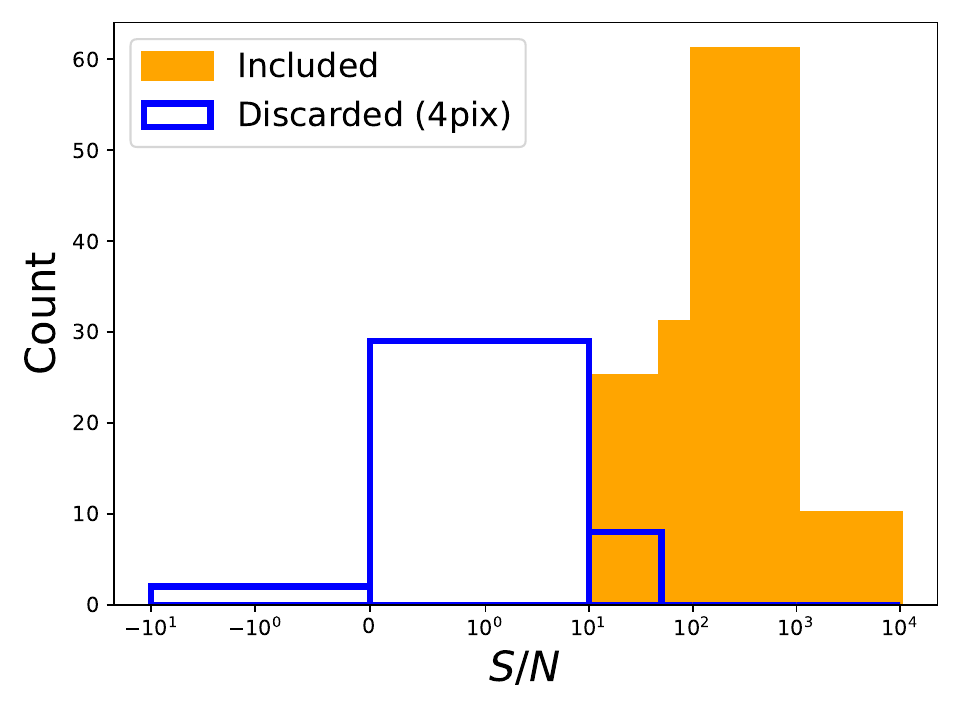}
    \caption{The $S/N$ distribution of the 113 galaxies from $HST$ imaging is shown with orange histograms. The $S/N$ distribution for the 39 discarded galaxies is shown with open blue histograms (with 4-pixel circular aperture) .} 
    \label{fig:snr_dist}
\end{figure}

In Fig.~\ref{fig:snr_dist}, we present the $S/N$ distribution of the galaxies used in this study. The solid orange histogram corresponds to the 113 galaxies included in our analysis, while the open blue histogram represents the 39 galaxies that were visually discarded. To estimate the $S/N$ of each object, we first derived the source flux (${\rm Flux}_{\rm src}$) in units of [${\rm electron}/s$] using the most probable magnitude ($m$) as:

\begin{equation}
{\rm Flux}_{\rm src} = 10^{-(m - m_0)/2.5},
\end{equation}
where $m_0$ is the zero-point magnitude for the corresponding filter. For the 39 visually discarded galaxies, we measured ${\rm Flux}_{\rm src}$ within a circular aperture of radius 4 pixels.

To account for correlated noise in the $HST$ images, we adopted an approach similar to that of \citet{Maulick_2024}. Specifically, we placed 1000 random apertures - each matching the size of the most probable {\tt GALFIT} model for the galaxy — in ``empty" regions of the $HST$ field. These regions were defined as those lying beyond a 50$\times$50 pixel box centered on any detected object, based on segmentation maps from {\sc SourceExtractor} with a $1.5\sigma$ detection threshold. The $S/N$ was then computed as the ratio of ${\rm Flux}_{\rm src}$ to the standard deviation of fluxes in these random apertures. All 113 selected galaxies have $S/N > 10$. Of the 39 discarded galaxies, 8 have $S/N \gtrsim 10$. These are primarily affected by contamination from nearby sources (e.g., HE0439-5254-127) or are extremely compact (unresolved), with an effective radius $R_e \sim 1$ pixel (e.g., HE0439-5254-244).

\section{Bayesian {\tt GALFIT} fitting procedure} 
\label{sec:fit-proc}

\begin{figure}
    \centering
    \includegraphics[width=0.8\linewidth]{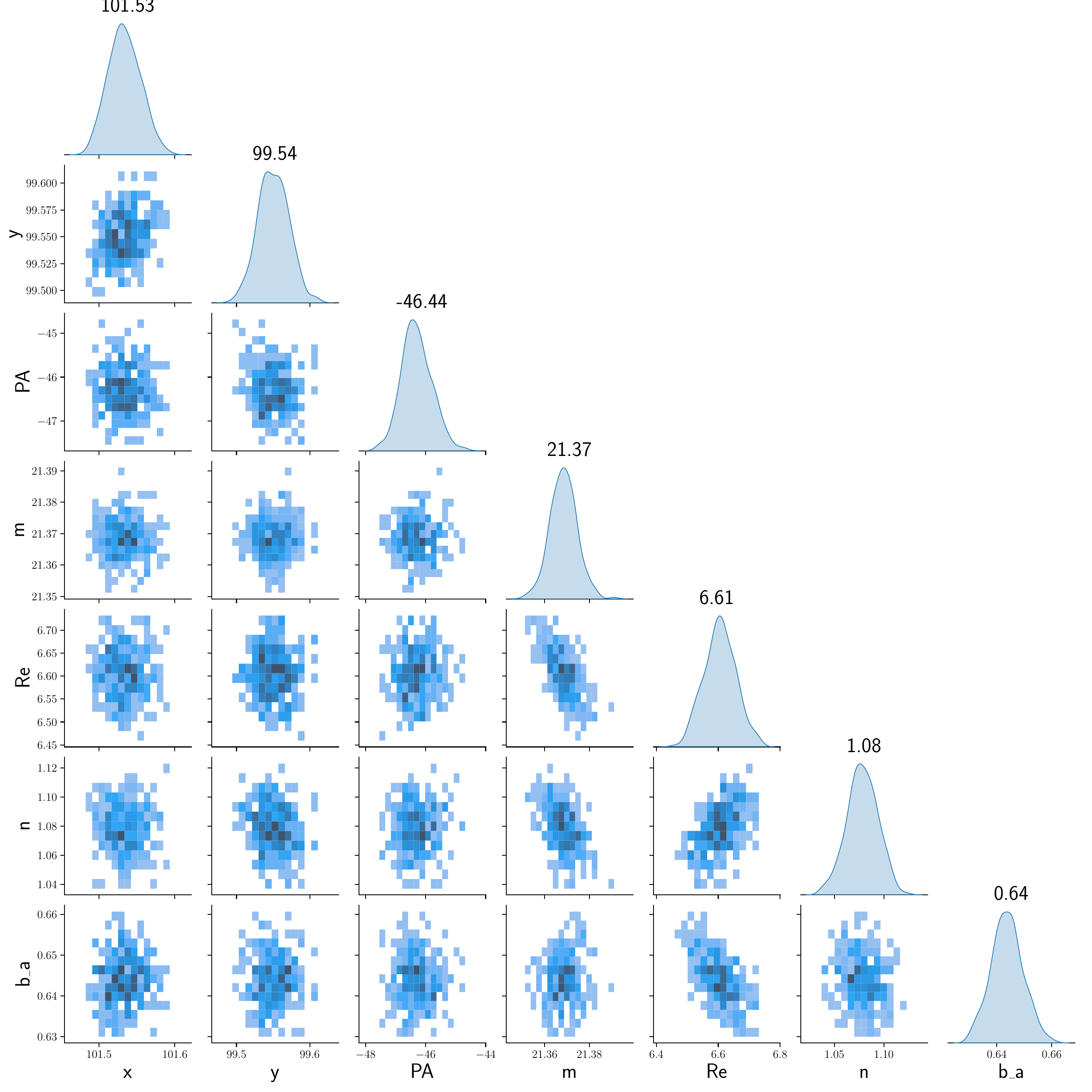}
    \includegraphics[width=1\linewidth]{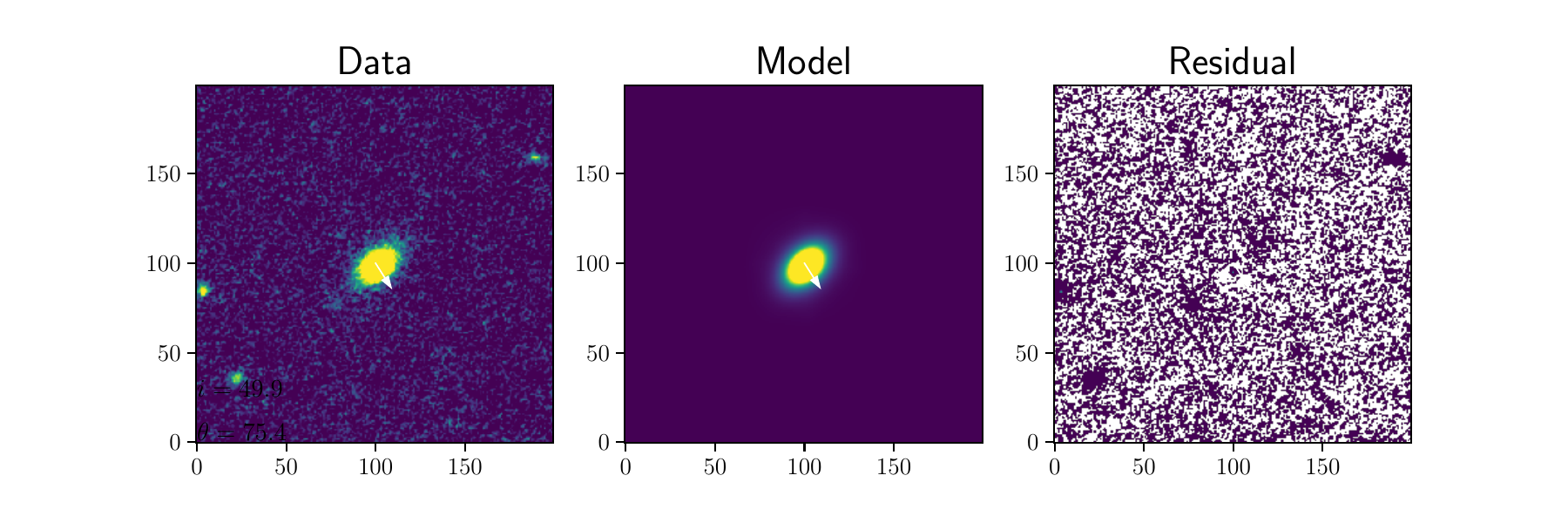}    
    \caption{Demonstration of the fitting procedure described in section \ref{sec:galfit}. The corner plot on top shows the posterior distribution of the 7 free parameters of the model, with the numbers on top indicating the most-probable values. The cutout, best-fit model, and the residuals are shown on the bottom panel. }
    \label{fig:demo-fit}
\end{figure}

In Fig.~\ref{fig:demo-fit}, we demonstrate the fitting procedure described in Sect \ref{sec:galfit}. 

\section{Distributions of the morphological parameters of the galaxies} 
\label{sec:best-fit-prop}
\begin{figure}
    \centering
    \includegraphics[width=1\linewidth]{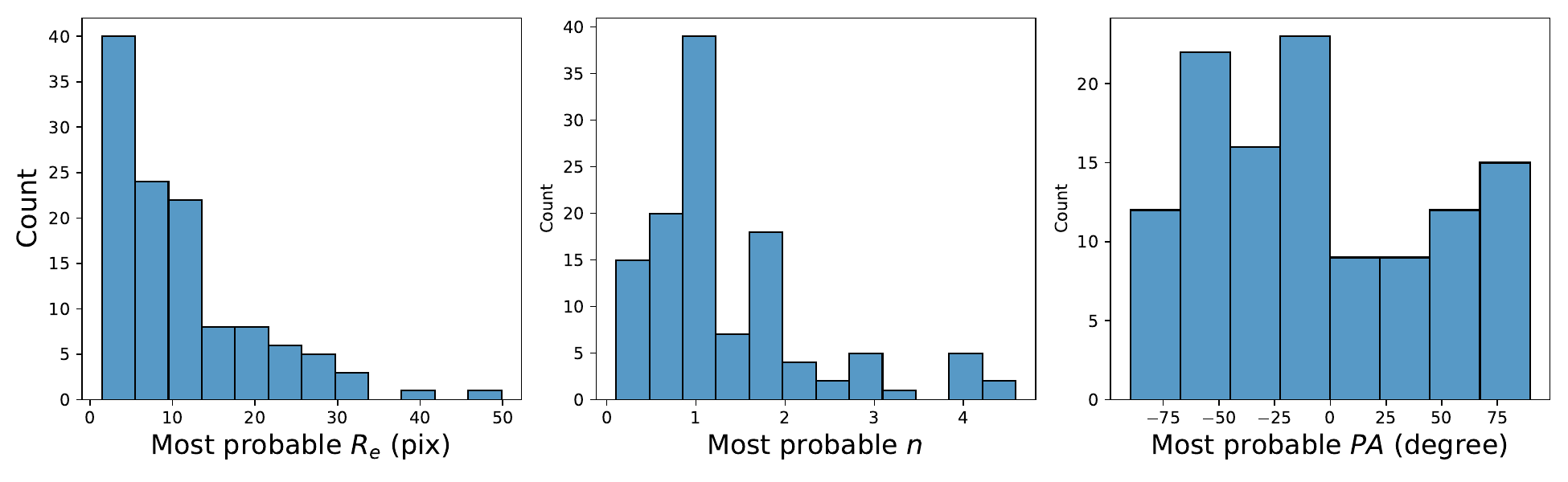}
    \caption{ Distribution of the most probable $R_e$, Sérsic index $n$, and position angle $PA$ are shown in the left, middle, and right panels. } 
    \label{fig:prop_dist}
\end{figure}

In the left, middle, and right panels of Fig.~\ref{fig:prop_dist}, we show the distribution of the most probable $R_e$, Sérsic index, and position angle. 

\section{Two $D/R_{\rm vir}$ for massive subsample} 

\begin{figure}
    \centering
    \includegraphics[width=0.3\linewidth]{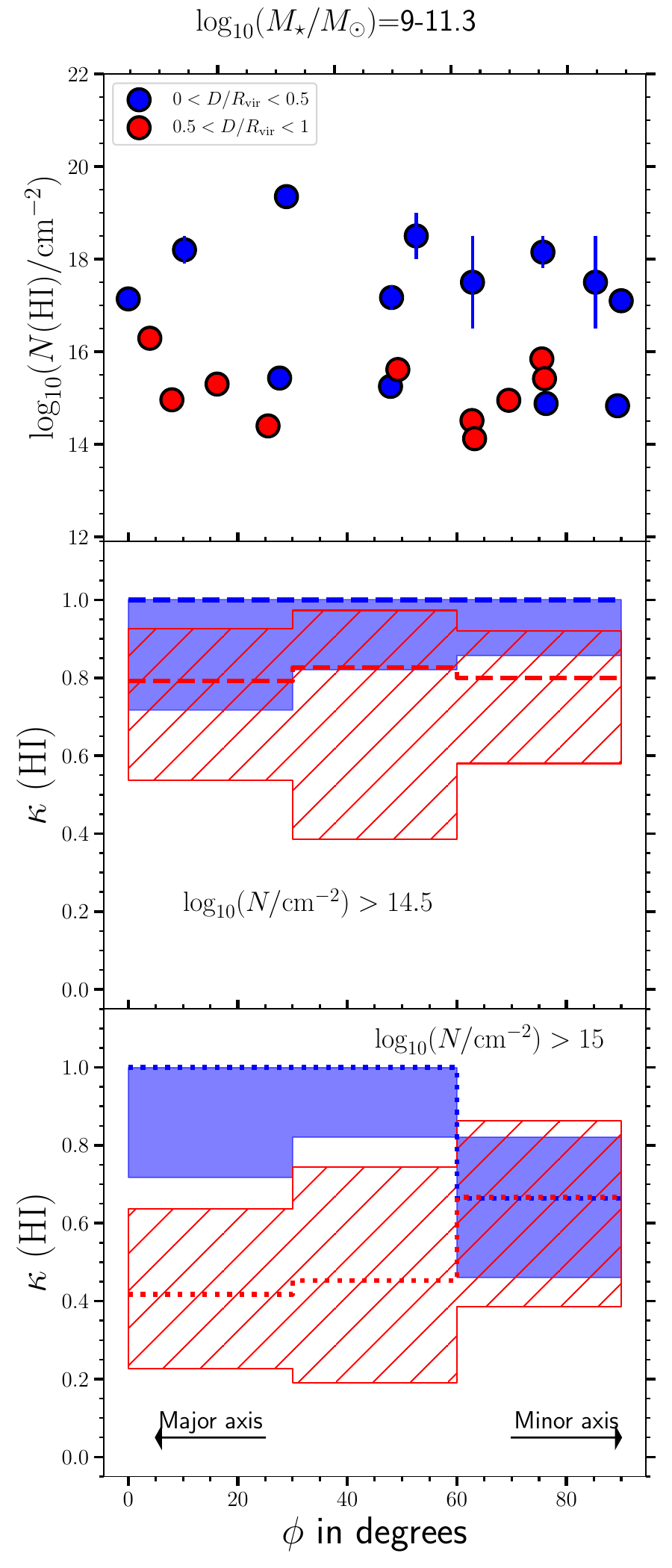}%
    \includegraphics[width=0.3\linewidth]{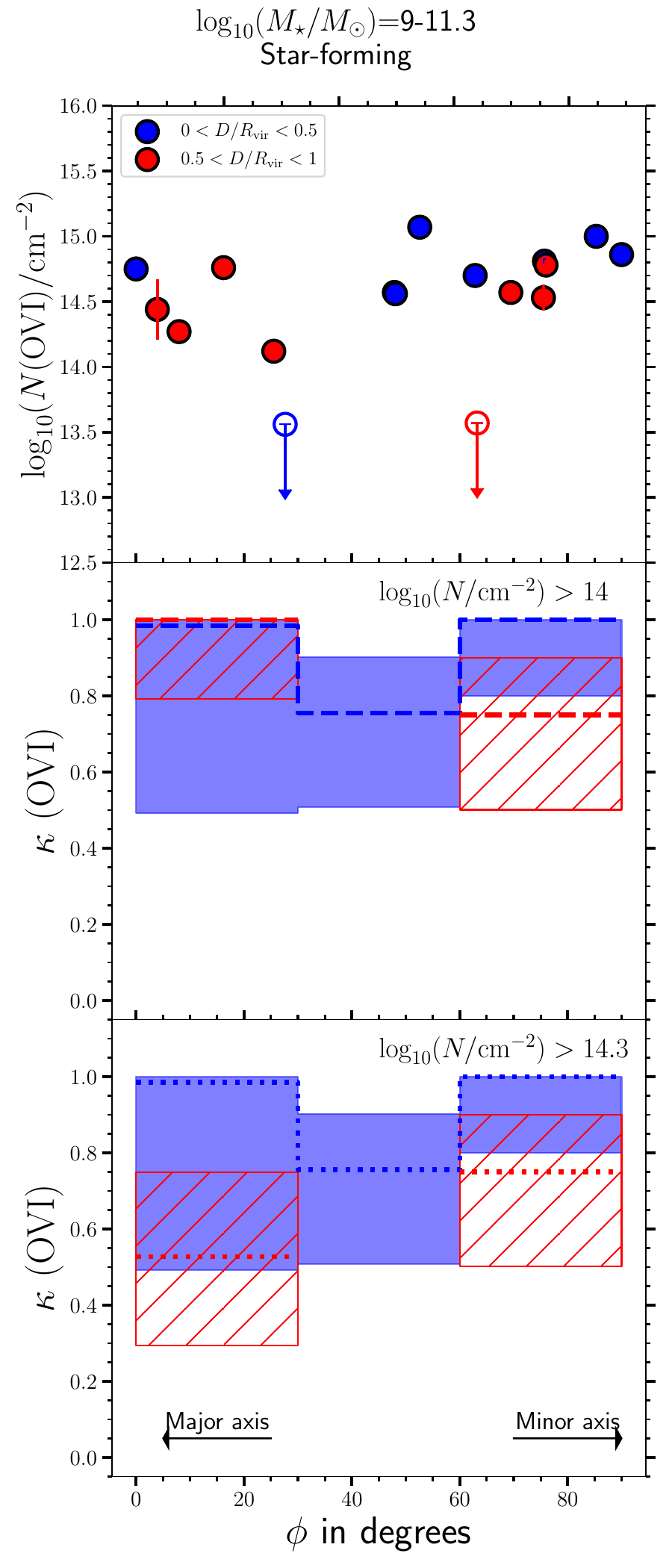}  
    \caption{Left and right panels are the same as the rightmost panels of Fig.~\ref{fig:hi_all} and Fig.~\ref{fig:ovi_all}, respectively, but split into $D/R_{\rm vir}=0-0.5$ and $0.5-1$ bins with blue and red colors (shown for edge-on galaxies only). The $\kappa(\HI)$ for $\phi=30^{\circ}-60^{\circ}$ is not shown in the right panel due to a lack of measurement of $D/R_{\rm vir}=0.5-1$ in this bin.}  
    \label{fig:app:2dn_bins}
\end{figure}

The edge-on ($i>35^{\circ}$), massive galaxy samples (massive and SF for the \OVI\ covering fraction analysis) within $R_{\rm vir}$ are further divided into two $D/R_{\rm vir}$ bins, with $D/R_{\rm vir}<0.5$ and $D/R_{\rm vir}=0.5-1$. The \kHI\ and \kOVI\ variation with azimuthal angle are shown in Fig.~\ref{fig:app:2dn_bins}. The lack of any significant azimuthal variation persists in both bins. 

\section{Variation of covering fraction beyond $R_{\rm vir}$}

\begin{figure}
    \centering
    \includegraphics[width=0.3\linewidth]{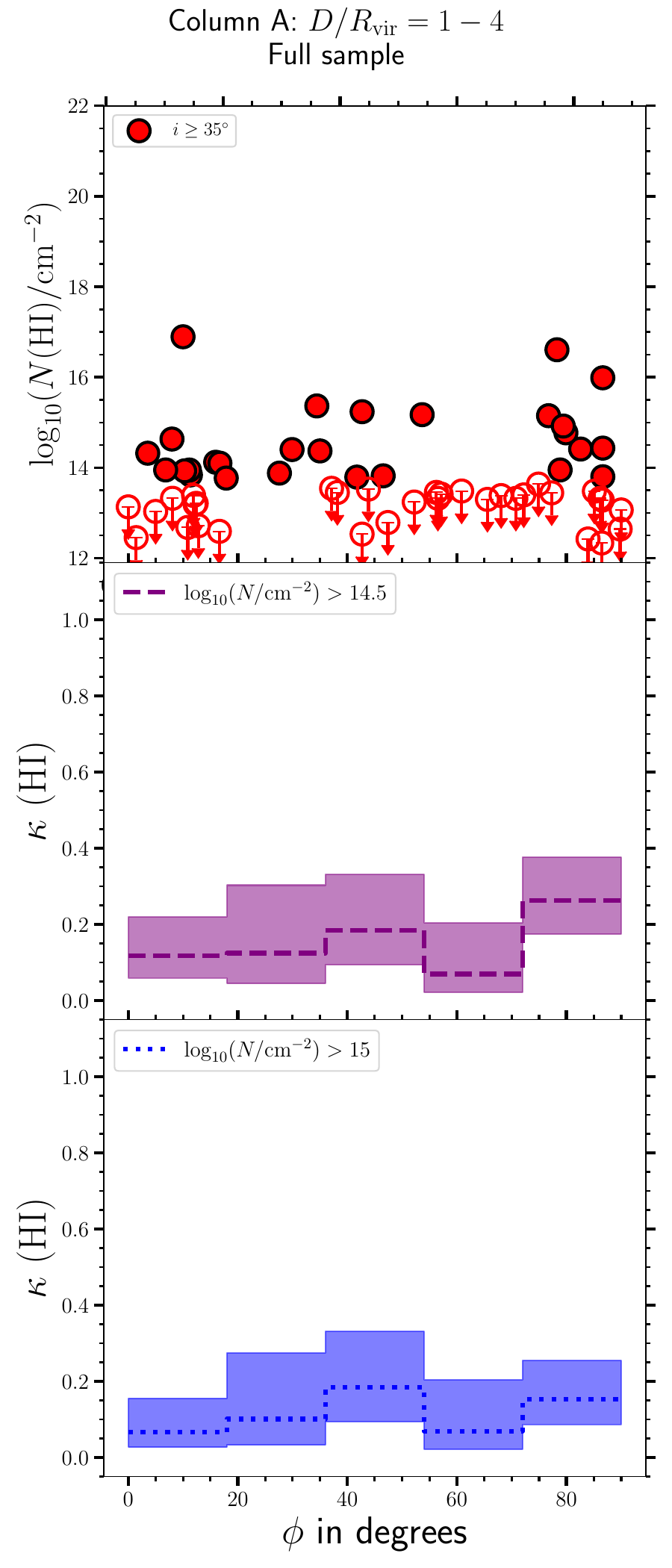}%
    \includegraphics[width=0.3\linewidth]{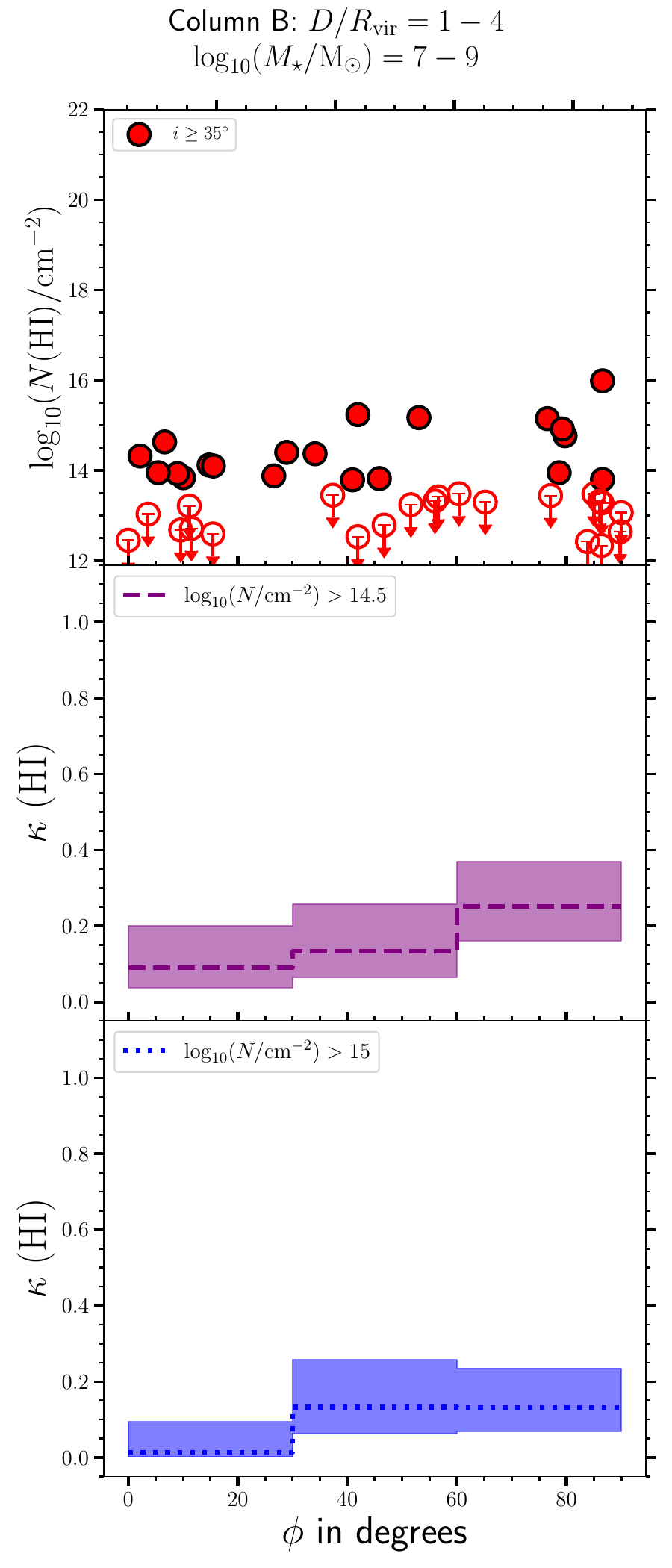}%
    \includegraphics[width=0.3\linewidth]{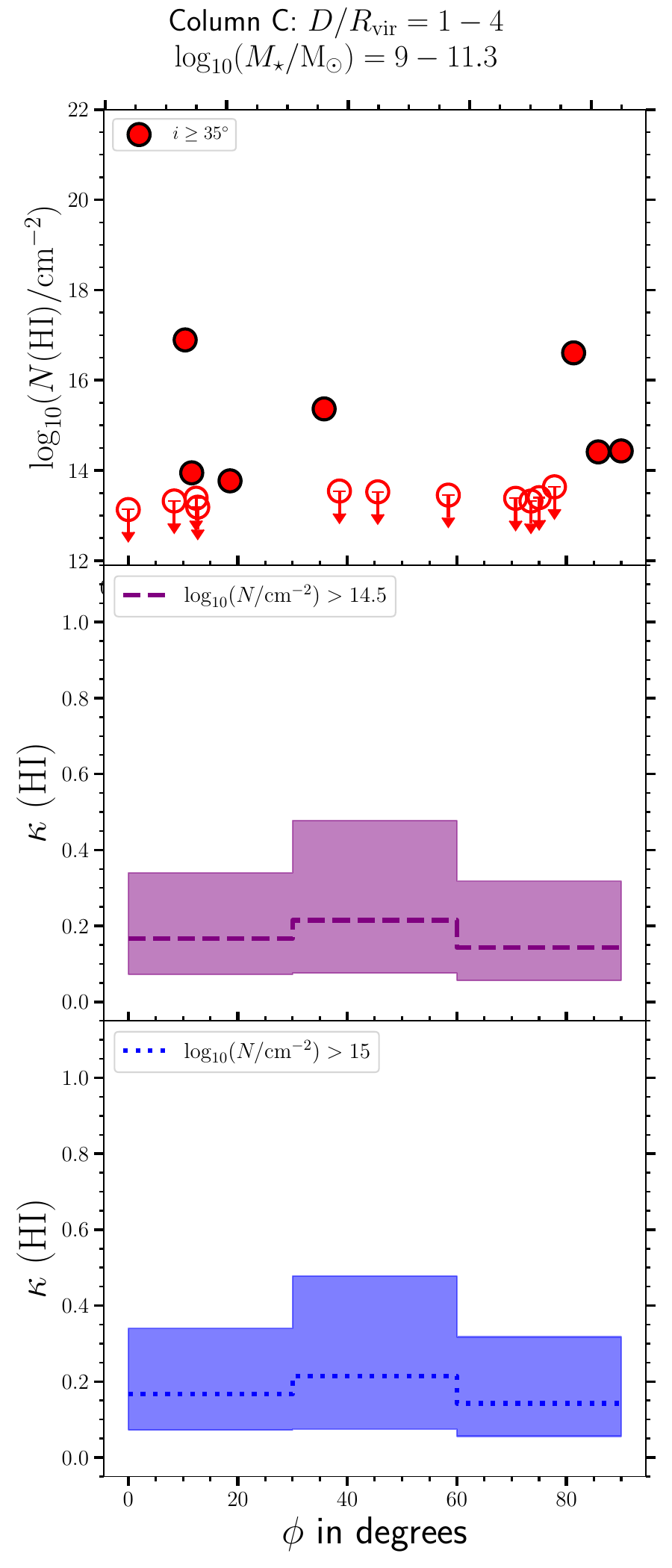}
    \caption{ Same as Fig.~\ref{fig:hi_all}, but for $D/R_{\rm vir}=1-4$, shown for edge-on galaxies only.} 
    \label{fig:app:out_vir}
\end{figure}

The \kHI\ variation with azimuthal angle for $D/R_{\rm vir}>1$ is shown in Fig.~\ref{fig:app:out_vir} for edge-on galaxies only. We find no statistically significant dependence of the \kHI\ on azimuthal angle beyond $R_{\rm vir}$.

The \kOVI\ beyond $R_{\rm vir}$ is substantially lower than that of \kHI, resulting in insufficient statistics to meaningfully assess its azimuthal dependence. We therefore do not present the corresponding \kOVI\ measurements here.

\section{Full Sample }
\label{sec:full_sample_table}

Table~\ref{tab:model_comparison} presents the full galaxy sample used in this work, along with the associated \HI\ and \OVI\ line measurements. 
\input{table_sample}

\bibliography{all_ref_morph}{}
\bibliographystyle{aasjournal}

\end{document}

%% file: table_sample.tex



\startlongtable
\begin{longrotatetable}
\begin{deluxetable*}{rrrrrrrrrrrrr}

\tablecaption{Summary of galaxy properties and \HI\ and \OVI\ line measurements used in this work}
\label{tab:model_comparison} 
\tablehead{
\colhead{\shortstack{$D$ in pkpc \\ (1)}} & \colhead{\shortstack{${\rm log}_{10}\left(\frac{M_{\star}}{{\rm M_{\odot}}}\right)$ \\ (2)}} & \colhead{\shortstack{$z$ \\ (3)}} & \colhead{\shortstack{$\phi_{\rm MP}$ \\ (4)}} & \colhead{\shortstack{$\phi[16\%]$ \\ (5)}} & \colhead{\shortstack{$\phi[84\%]$ \\ (6)}} & \colhead{\shortstack{$i_{\rm MP}$ \\ (7)}} & \colhead{\shortstack{$i[16\%]$ \\ (8)}} & \colhead{\shortstack{$i[84\%]$ \\ (9)}} & \colhead{\shortstack{${\rm log}_{10}\left(\frac{N_{\rm HI}}{{\rm cm}^{-2}}\right)$ \\ (10)}} & \colhead{\shortstack{d$\left[{\rm log}_{10}\left(\frac{N_{\rm HI}}{{\rm cm}^{-2}}\right)\right]$ \\ (11)}} & \colhead{\shortstack{${\rm log}_{10}\left(\frac{N_{\rm OVI}}{{\rm cm}^{-2}}\right)$ \\ (12)}} & \colhead{\shortstack{d$\left[{\rm log}_{10}\left(\frac{N_{\rm OVI}}{{\rm cm}^{-2}}\right)\right]$ \\ (13)}}
}


\startdata
171.2 & 8.2 & 0.5099 & 56.76 & 51.01 & 63.40 & 50.90 & 53.09 & 43.33 & 13.32 & -999.00 & 13.47 & -999.00 \\
34.5 & 8.4 & 0.3329 & 8.74 & 4.84 & 12.14 & 55.95 & 58.40 & 51.98 & 14.14 & 0.02 & 13.60 & 0.05 \\
43.1 & 8.7 & 0.3250 & 80.18 & 79.59 & 80.77 & 74.78 & 75.77 & 73.05 & 16.48 & 0.02 & 14.23 & 0.27 \\
87.8 & 9.8 & 0.3281 & 75.59 & 75.13 & 75.78 & 63.24 & 63.50 & 62.97 & 15.84 & 0.02 & 14.53 & 0.10 \\
51.2 & 8.4 & 0.3835 & 74.86 & 70.76 & 79.61 & 40.72 & 42.93 & 36.81 & 14.80 & 0.01 & 14.10 & 0.03 \\
71.2 & 8.7 & 0.2493 & 71.98 & 69.14 & 73.39 & 34.87 & 36.95 & 32.91 & 15.71 & 0.01 & 13.61 & 0.05 \\
180.9 & 8.3 & 0.5319 & 36.58 & 34.58 & 39.70 & 64.86 & 66.67 & 62.52 & 14.37 & 0.03 & 13.87 & 0.05 \\
80.8 & 8.2 & 0.4019 & 11.26 & 10.22 & 17.83 & 56.22 & 60.10 & 51.93 & 14.63 & 0.03 & 13.48 & -999.00 \\
70.8 & 8.2 & 0.2108 & 39.46 & 38.32 & 40.54 & 75.05 & 75.94 & 74.04 & 12.44 & -999.00 & 13.47 & -999.00 \\
158.1 & 8.1 & 0.3995 & 55.95 & 46.83 & 57.20 & 28.20 & 31.26 & 26.01 & 13.66 & 0.04 & 13.48 & -999.00 \\
220.4 & 9.2 & 0.5941 & 67.57 & 66.42 & 68.74 & 72.70 & 73.86 & 71.27 & 13.38 & -999.00 & 13.56 & -999.00 \\
158.2 & 7.9 & 0.2653 & 18.74 & 6.50 & 27.60 & 44.77 & 50.36 & 35.02 & 14.12 & 0.02 & 13.98 & -999.00 \\
180.1 & 10.1 & 0.4319 & 81.17 & 80.95 & 81.35 & 64.14 & 64.31 & 63.94 & 14.41 & 0.02 & 13.98 & 0.10 \\
147.7 & 8.8 & 0.3226 & 63.15 & 62.31 & 69.39 & 27.03 & 28.79 & 24.03 & 16.48 & 0.02 & 13.38 & 0.07 \\
179.8 & 8.2 & 0.4791 & 65.22 & 63.75 & 66.65 & 59.55 & 60.83 & 58.18 & 13.30 & -999.00 & 13.75 & -999.00 \\
210.4 & 8.3 & 0.4225 & 19.37 & 14.25 & 23.11 & 52.79 & 56.24 & 49.66 & 12.59 & -999.00 & 13.50 & -999.00 \\
131.0 & 8.3 & 0.4161 & 19.46 & 14.59 & 21.97 & 69.91 & 72.48 & 65.59 & 14.10 & 0.03 & 13.42 & -999.00 \\
87.8 & 7.0 & 0.2175 & 2.79 & 1.80 & 10.09 & 52.79 & 58.68 & 47.12 & -999.00 & -999.00 & -999.00 & -999.00 \\
135.2 & 9.5 & 0.5591 & 56.49 & 55.31 & 58.96 & 70.54 & 72.32 & 68.76 & 13.45 & -999.00 & 13.76 & -999.00 \\
236.9 & 10.1 & 0.7274 & 44.87 & 43.57 & 45.90 & 45.86 & 46.54 & 44.93 & 13.53 & -999.00 & 13.79 & -999.00 \\
79.7 & 7.9 & 0.2848 & 84.95 & 76.28 & 87.34 & 67.39 & 75.98 & 60.78 & 13.80 & 0.08 & 13.51 & -999.00 \\
196.4 & 8.4 & 0.5738 & 83.51 & 82.12 & 84.70 & 82.34 & 84.92 & 80.91 & 13.48 & -999.00 & 13.73 & -999.00 \\
144.0 & 7.3 & 0.3949 & 29.64 & 20.30 & 36.68 & 80.45 & 85.33 & 71.82 & 13.88 & 0.03 & 13.85 & -999.00 \\
190.7 & 7.9 & 0.6830 & 39.55 & 36.87 & 44.12 & 75.68 & 83.63 & 67.97 & 13.45 & -999.00 & 13.67 & -999.00 \\
99.0 & 8.6 & 0.2019 & 43.78 & 43.02 & 44.73 & 56.85 & 57.54 & 55.82 & 15.24 & 0.04 & 14.00 & 0.05 \\
221.9 & 7.9 & 0.4355 & 14.41 & 11.59 & 63.62 & 35.49 & 44.53 & 19.13 & 13.84 & 0.05 & 13.66 & -999.00 \\
196.7 & 10.1 & 0.3991 & 13.15 & 12.83 & 15.40 & 69.37 & 72.09 & 67.68 & 16.89 & 0.06 & 14.39 & 0.02 \\
94.2 & 9.6 & 0.5172 & 34.14 & 33.30 & 37.36 & 26.13 & 27.17 & 24.04 & 14.60 & 0.12 & 14.47 & 0.05 \\
104.2 & 8.8 & 0.2606 & 78.65 & 78.42 & 79.29 & 66.67 & 67.19 & 66.36 & 14.77 & 0.02 & -999.00 & -999.00 \\
251.7 & 10.1 & 0.7369 & 51.80 & 46.03 & 55.73 & 32.52 & 35.10 & 30.55 & 13.73 & -999.00 & 14.01 & -999.00 \\
244.6 & 9.2 & 0.3994 & 73.96 & 73.77 & 74.18 & 75.68 & 76.60 & 74.61 & 13.64 & -999.00 & 13.71 & -999.00 \\
189.0 & 8.7 & 0.1775 & 84.86 & 84.39 & 85.53 & 49.73 & 50.56 & 49.38 & 13.28 & -999.00 & -999.00 & -999.00 \\
194.1 & 8.3 & 0.4339 & 42.88 & 42.54 & 43.42 & 69.64 & 70.02 & 69.24 & 13.79 & 0.06 & 13.91 & -999.00 \\
68.1 & 9.2 & 0.1732 & 63.60 & 63.19 & 63.99 & 45.05 & 45.27 & 44.72 & 14.51 & 0.04 & -999.00 & -999.00 \\
139.5 & 9.3 & 0.7315 & 38.56 & 29.28 & 46.22 & 53.87 & 60.82 & 47.01 & 13.54 & -999.00 & 13.80 & -999.00 \\
163.0 & 8.6 & 0.5554 & 60.81 & 57.60 & 63.98 & 47.66 & 49.81 & 45.00 & 13.48 & -999.00 & 13.89 & -999.00 \\
74.1 & 6.1 & 0.1819 & 25.68 & 25.09 & 29.23 & 87.12 & 87.87 & 85.70 & -999.00 & -999.00 & -999.00 & -999.00 \\
46.6 & 8.5 & 0.2958 & 31.26 & 30.13 & 31.84 & 71.71 & 72.33 & 70.75 & 14.14 & 0.02 & 13.53 & -999.00 \\
13.6 & 7.1 & 0.1317 & 61.71 & 58.47 & 65.05 & 67.12 & 70.85 & 64.15 & 14.52 & 0.10 & 13.39 & -999.00 \\
103.7 & 7.6 & 0.2252 & 51.35 & 29.97 & 68.53 & 31.80 & 38.22 & 18.77 & 16.98 & 0.30 & 14.20 & 0.01 \\
72.7 & 7.9 & 0.2906 & 31.80 & 30.20 & 33.81 & 54.87 & 56.57 & 53.59 & 14.40 & 0.11 & 14.11 & 0.04 \\
67.6 & 8.7 & 0.4002 & 29.82 & 28.08 & 30.90 & 62.52 & 63.78 & 61.23 & 14.61 & 0.02 & 14.32 & 0.02 \\
95.4 & 7.2 & 0.2218 & 75.68 & 73.62 & 76.88 & 64.78 & 65.98 & 63.10 & 15.15 & 0.02 & 13.83 & -999.00 \\
160.7 & 8.7 & 0.4193 & 5.13 & 2.75 & 9.25 & 47.66 & 50.64 & 44.26 & 12.45 & -999.00 & 13.68 & -999.00 \\
116.4 & 8.4 & 0.3090 & 7.12 & 6.44 & 7.67 & 73.51 & 74.07 & 72.88 & 14.32 & 0.01 & 13.16 & -999.00 \\
183.1 & 8.4 & 0.3726 & 39.55 & 34.70 & 43.50 & 67.66 & 71.40 & 62.00 & -999.00 & -999.00 & 13.29 & -999.00 \\
169.0 & 7.6 & 0.5320 & 52.70 & 43.01 & 67.39 & 56.67 & 63.51 & 45.11 & 13.24 & -999.00 & 13.43 & -999.00 \\
149.9 & 9.5 & 0.6016 & 20.54 & 17.01 & 24.20 & 63.96 & 67.94 & 61.13 & 13.77 & 0.06 & 13.59 & -999.00 \\
69.4 & 9.4 & 0.4243 & 69.91 & 69.61 & 70.33 & 72.61 & 73.14 & 72.46 & 14.95 & 0.10 & 14.57 & 0.01 \\
28.8 & 7.6 & 0.5723 & 71.98 & 65.02 & 76.27 & 58.65 & 63.50 & 53.15 & 14.98 & 0.02 & 14.31 & 0.03 \\
139.5 & 7.8 & 0.3000 & 15.77 & 9.12 & 26.80 & 37.30 & 45.65 & 30.43 & 12.72 & -999.00 & 13.61 & -999.00 \\
70.6 & 8.9 & 0.5160 & 32.61 & 30.68 & 35.26 & 73.15 & 73.91 & 72.25 & 13.43 & -999.00 & 13.65 & -999.00 \\
167.1 & 9.2 & 0.6490 & 3.15 & 2.69 & 15.79 & 26.85 & 28.73 & 23.22 & 13.49 & -999.00 & 13.68 & -999.00 \\
179.8 & 8.6 & 0.2019 & 88.11 & 87.74 & 89.37 & 56.67 & 57.56 & 55.88 & 13.07 & -999.00 & 13.59 & -999.00 \\
273.4 & 10.1 & 0.3991 & 70.09 & 68.30 & 71.71 & 73.42 & 75.06 & 71.44 & 13.32 & -999.00 & 13.63 & -999.00 \\
298.5 & 8.8 & 0.2606 & 8.47 & 8.25 & 9.17 & 66.85 & 67.21 & 66.34 & 13.03 & -999.00 & -999.00 & -999.00 \\
177.8 & 9.2 & 0.3994 & 84.95 & 84.75 & 85.14 & 75.77 & 75.92 & 75.56 & 14.43 & 0.05 & 14.38 & 0.05 \\
138.7 & 9.1 & 0.4879 & 71.44 & 70.41 & 72.26 & 73.69 & 75.73 & 71.31 & 13.40 & -999.00 & 13.76 & -999.00 \\
43.8 & 8.7 & 0.1775 & 75.41 & 74.75 & 75.79 & 49.91 & 50.37 & 49.47 & 14.80 & 0.05 & -999.00 & -999.00 \\
40.8 & 9.3 & 0.3846 & 88.56 & 86.46 & 89.28 & 72.34 & 74.22 & 70.81 & 14.83 & 0.03 & 14.37 & 0.08 \\
72.1 & 9.1 & 0.4298 & 28.65 & 26.81 & 30.29 & 58.11 & 58.95 & 56.51 & 14.40 & 0.05 & 14.12 & 0.03 \\
285.8 & 6.8 & 0.5399 & 47.93 & 46.93 & 51.64 & 58.83 & 59.77 & 57.64 & 13.15 & -999.00 & 13.39 & -999.00 \\
225.6 & 9.4 & 0.6614 & 15.22 & 14.41 & 16.42 & 78.47 & 79.47 & 77.60 & 13.19 & -999.00 & 13.39 & -999.00 \\
236.3 & 8.8 & 0.7149 & 15.40 & 13.03 & 18.04 & 78.29 & 79.93 & 76.24 & 13.21 & -999.00 & 13.39 & -999.00 \\
69.0 & 8.8 & 0.1838 & 60.00 & 58.98 & 60.82 & 41.98 & 42.74 & 41.54 & -999.00 & -999.00 & 14.05 & 0.03 \\
134.0 & 9.3 & 0.3002 & 36.04 & 28.09 & 46.83 & 55.86 & 60.83 & 54.86 & 15.37 & 0.04 & 13.41 & -999.00 \\
195.3 & 8.9 & 0.4787 & 13.96 & 12.13 & 15.12 & 55.95 & 56.91 & 54.61 & 12.68 & -999.00 & 13.67 & -999.00 \\
207.5 & 10.3 & 0.5356 & 78.74 & 71.28 & 81.51 & 26.58 & 29.37 & 23.18 & 14.38 & 0.02 & 13.62 & -999.00 \\
173.4 & 9.0 & 0.4581 & 73.51 & 70.00 & 79.90 & 28.29 & 29.73 & 25.91 & 12.58 & -999.00 & 13.46 & -999.00 \\
82.8 & 7.9 & 0.1390 & 82.43 & 80.77 & 85.29 & 49.37 & 51.13 & 47.80 & 12.42 & -999.00 & 13.61 & -999.00 \\
126.6 & 8.7 & 0.6531 & 84.50 & 75.66 & 86.75 & 40.18 & 43.05 & 36.48 & 13.32 & -999.00 & 13.56 & -999.00 \\
103.7 & 10.2 & 0.5722 & 50.90 & 49.12 & 54.24 & 48.20 & 50.51 & 47.55 & 15.61 & 0.08 & 13.76 & -999.00 \\
178.2 & 9.5 & 0.7293 & 77.12 & 76.58 & 77.46 & 63.78 & 64.43 & 63.49 & 16.61 & 0.04 & 13.64 & -999.00 \\
50.5 & 8.5 & 0.3199 & 73.78 & 73.25 & 74.21 & 75.05 & 75.65 & 74.58 & 14.55 & 0.01 & 13.66 & -999.00 \\
170.4 & 9.6 & 0.5206 & 11.35 & 11.09 & 12.06 & 60.09 & 60.54 & 59.70 & 13.32 & -999.00 & 13.58 & -999.00 \\
212.0 & 8.9 & 0.4396 & 43.78 & 43.57 & 44.65 & 54.41 & 55.84 & 52.79 & 12.53 & -999.00 & 13.60 & -999.00 \\
193.4 & 7.8 & 0.3613 & 54.05 & 32.31 & 65.23 & 36.76 & 43.24 & 25.59 & 15.17 & 0.01 & 13.87 & 0.01 \\
149.5 & 8.1 & 0.4831 & 77.66 & 73.03 & 79.89 & 67.93 & 69.78 & 60.80 & 13.95 & 0.02 & 13.35 & -999.00 \\
48.0 & 7.9 & 0.1823 & 6.40 & 5.53 & 6.91 & 70.09 & 70.82 & 69.47 & 14.83 & 0.07 & 14.03 & 0.06 \\
42.1 & 7.5 & 0.1485 & 17.84 & 15.73 & 20.26 & 57.30 & 59.47 & 55.25 & 15.16 & 0.04 & 13.51 & -999.00 \\
165.2 & 9.4 & 0.4249 & 14.23 & 12.57 & 14.77 & 66.22 & 66.63 & 65.83 & 13.95 & 0.03 & 13.42 & -999.00 \\
95.3 & 8.7 & 0.3768 & 13.42 & 13.17 & 13.95 & 72.88 & 73.19 & 72.44 & 13.93 & 0.03 & 13.38 & -999.00 \\
72.8 & 7.7 & 0.1950 & 84.78 & 84.11 & 85.23 & 75.95 & 76.38 & 75.60 & 12.34 & -999.00 & 13.78 & -999.00 \\
63.0 & 6.9 & 0.4471 & 84.95 & 67.02 & 86.25 & 63.78 & 68.61 & 58.67 & 15.99 & 0.04 & 13.77 & 0.06 \\
108.1 & 9.8 & 0.2443 & 20.36 & 19.44 & 21.34 & 34.77 & 35.65 & 34.48 & 15.45 & 0.03 & 14.69 & 0.04 \\
95.7 & 10.8 & 0.1540 & 31.80 & 23.18 & 32.86 & 39.19 & 42.05 & 37.25 & 19.35 & 0.15 & 14.19 & -999.00 \\
119.6 & 10.8 & 0.2284 & 30.63 & 30.32 & 31.07 & 51.35 & 51.84 & 51.25 & 15.43 & 0.08 & 13.56 & -999.00 \\
155.8 & 11.0 & 0.1431 & 76.31 & 71.44 & 76.87 & 64.59 & 65.34 & 64.13 & 14.88 & 0.06 & 13.80 & -999.00 \\
100.2 & 9.6 & 0.3529 & 8.38 & 7.07 & 15.88 & 39.19 & 40.18 & 36.94 & 16.29 & 0.03 & 14.44 & 0.23 \\
93.7 & 11.2 & 0.2119 & 14.32 & 14.13 & 14.59 & 48.20 & 48.84 & 48.09 & 18.20 & 0.30 & 14.28 & 0.04 \\
215.5 & 9.5 & 0.5100 & 14.96 & 14.03 & 16.59 & 47.84 & 48.81 & 46.89 & 13.39 & -999.00 & 13.54 & -999.00 \\
159.8 & 10.6 & 0.4772 & 64.59 & 10.02 & 72.10 & 5.04 & 13.04 & 3.44 & 12.81 & -999.00 & 13.57 & -999.00 \\
123.0 & 9.7 & 0.3721 & 64.05 & 63.63 & 64.43 & 73.60 & 73.80 & 72.31 & 14.12 & 0.03 & 13.57 & -999.00 \\
132.1 & 8.1 & 0.4400 & 87.93 & 87.22 & 88.18 & 74.05 & 74.80 & 73.84 & 12.64 & -999.00 & 13.67 & -999.00 \\
158.6 & 8.1 & 0.7105 & 76.22 & 73.98 & 78.23 & 66.58 & 68.95 & 64.36 & 13.44 & -999.00 & 13.62 & -999.00 \\
65.8 & 9.9 & 0.2278 & 84.78 & 83.42 & 85.42 & 47.66 & 48.28 & 44.27 & 17.50 & 1.00 & 15.00 & 0.03 \\
46.1 & 10.3 & 0.3557 & 54.05 & 52.55 & 57.85 & 39.10 & 39.84 & 38.10 & 18.50 & 0.50 & 15.07 & 0.02 \\
21.8 & 10.2 & 0.2520 & 89.19 & 88.64 & 89.63 & 52.61 & 52.95 & 52.18 & 17.10 & 0.02 & 14.86 & 0.02 \\
46.1 & 10.5 & 0.1661 & 63.69 & 59.84 & 65.20 & 42.52 & 43.48 & 34.21 & 17.50 & 1.00 & 14.70 & 0.03 \\
54.2 & 10.3 & 0.2467 & 79.28 & 69.40 & 81.78 & 29.91 & 35.79 & 26.11 & 16.70 & 0.90 & 14.70 & 0.04 \\
14.9 & 11.2 & 0.2367 & 23.87 & 18.19 & 24.62 & 34.05 & 37.55 & 33.57 & 18.60 & 0.06 & 14.07 & -999.00 \\
55.2 & 10.1 & 0.1545 & 77.39 & 76.42 & 79.42 & 26.04 & 26.91 & 24.20 & 16.90 & 1.10 & 14.43 & 0.08 \\
29.7 & 10.6 & 0.3185 & 34.87 & 32.38 & 37.43 & 32.97 & 34.58 & 29.98 & 15.57 & 0.02 & 14.68 & 0.03 \\
21.8 & 10.2 & 0.2053 & 75.77 & 75.46 & 76.10 & 53.87 & 54.82 & 53.68 & 18.15 & 0.35 & 14.81 & 0.02 \\
95.1 & 10.1 & 0.2178 & 19.91 & 19.07 & 20.32 & 40.27 & 40.59 & 39.78 & 15.30 & 0.06 & 14.76 & 0.02 \\
37.8 & 10.8 & 0.2142 & 4.68 & 4.39 & 7.53 & 53.60 & 53.97 & 52.37 & 17.14 & 0.04 & 14.75 & 0.02 \\
248.5 & 8.4 & 0.4950 & 57.30 & 56.03 & 58.91 & 72.43 & 74.09 & 71.09 & 13.42 & -999.00 & 13.65 & -999.00 \\
257.4 & 9.7 & 0.4778 & 3.78 & 3.32 & 4.27 & 59.91 & 60.36 & 59.44 & 13.13 & -999.00 & 13.73 & -999.00 \\
116.4 & 8.6 & 0.2158 & 10.18 & 9.43 & 10.69 & 70.27 & 70.79 & 69.75 & 13.95 & 0.02 & 13.57 & -999.00 \\
112.7 & 8.9 & 0.3567 & 48.20 & 46.27 & 49.41 & 62.61 & 63.63 & 60.70 & 12.79 & -999.00 & 13.72 & -999.00 \\
117.8 & 8.0 & 0.3986 & 47.39 & 42.68 & 52.47 & 60.90 & 63.79 & 56.32 & 13.82 & 0.05 & 13.49 & -999.00 \\
100.4 & 9.4 & 0.6104 & 76.04 & 74.02 & 81.19 & 38.92 & 40.06 & 35.78 & 15.42 & 0.02 & 14.78 & 0.04 \\
60.4 & 7.4 & 0.4436 & 78.20 & 76.29 & 80.88 & 68.02 & 69.91 & 65.81 & 14.92 & 0.19 & 13.60 & -999.00 \\
88.7 & 10.6 & 0.1792 & 13.87 & 6.00 & 15.10 & 17.48 & 23.10 & 16.82 & 16.63 & 0.30 & 14.59 & 0.05 \\
85.1 & 10.4 & 0.2623 & 49.64 & 49.48 & 50.03 & 56.76 & 57.02 & 56.59 & 15.25 & 0.06 & 14.57 & 0.05 \\
38.7 & 11.1 & 0.2024 & 76.85 & 72.46 & 82.88 & 32.70 & 34.77 & 30.04 & 19.80 & 0.10 & 14.73 & 0.04 \\
35.9 & 10.5 & 0.1893 & 49.82 & 49.73 & 50.04 & 53.24 & 53.37 & 53.10 & 17.17 & 0.26 & 14.56 & 0.05 \\
99.6 & 10.1 & 0.1414 & 12.16 & 12.10 & 12.32 & 73.69 & 73.83 & 73.65 & 14.96 & 0.03 & 14.27 & 0.05 \\
\enddata

\tablecomments{(1) Impact parameter (2) Stellar mass of the galaxy (3) Galaxy redshift (4) Most probable azimuthal angle (5) 16$^{\rm th}$ percentile of the posterior azimuthal angle distribution (6) 84$^{\rm th}$ percentile of the posterior azimuthal angle distribution (7)  Most probable inclination angle (8) 16$^{\rm th}$ percentile of the posterior inclination angle distribution (9) 84$^{\rm th}$ percentile of the posterior inclination angle distribution (10) Total \HI\ column density (Flag value of -999.0 indicate unavailability of the measurement) or the upper limit on the total \HI\ column density when column [12] has a flag value of -999.0 (12) Uncertainty on the total \HI\ column density. Flag value of -999.0 either indicate non-detection of \HI\ (when column [11] is not flagged) or unavailability of the measurement (when column [11] is flagged) (13) Total \OVI\ column density (Flag value of -999.0 indicate unavailability of the measurement) or the upper limit on the total \OVI\ column density when column [14] has a flag value of -999.0 (14) Uncertainty on the total \OVI\ column density. Flag value of -999.0 either indicate non-detection of \OVI\ (when column [13] is not flagged) or unavailability of the measurement (when column [13] is flagged)} 
\end{deluxetable*}
\end{longrotatetable}

